\documentclass[draft, jgrga]{AGUTeX}

\usepackage{lineno}
\linenumbers*[1]

  \usepackage{graphicx}  
  \usepackage{amssymb}
\usepackage{amsmath}
\usepackage{url}
 \setkeys{Gin}{draft=false}
\newcommand{\abs}[1]{\left |#1\right |}

\newcommand{\paren}[1]{\left(#1\right)}
\newcommand{\bracket}[1]{\left[#1\right]}
\newcommand{\set}[1]{\left\{#1\right\}}

\newcommand{\mean}[1]{\langle#1\rangle}
\newcommand{\eps}{\epsilon}

\newcommand{\comp}{\mathrm{comp.}}

\newcommand{\inc}{\mathrm{inc.}}

\newcommand{\f}{\phi}

\newcommand{\R}{\mathbb{R}}

\newcommand{\by}{\mathbf{y}}
\newcommand{\bv}{\mathbf{v}}
\newcommand{\bV}{\mathbf{V}}

\newcommand{\bg}{\mathbf{g}}
\newcommand{\bk}{\mathbf{k}}

\newcommand{\bn}{\mathbf{n}}

\newcommand{\eff}{{\textrm{eff.}}}

\authorrunninghead{SIMPSON ET AL.}

\titlerunninghead{A MULTISCALE MODEL OF PARTIAL MELTS}

\authoraddr{G. Simpson,
Department of Mathematics, University of Toronto, Toronto, ON M5S 2E4, Canada.
(simpson@math.toronto.edu)}

\authoraddr{M. Spiegelman,
Department of Applied Physics and Applied Mathematics, Columbia University, New York, NY 10027, USA.
Lamont-Doherty Earth Observatory, Palisades, NY 10964, USA.
(mspieg@ldeo.columbia.edu)}

\authoraddr{M. I. Weinstein,
Department of Applied Physics and Applied Mathematics, Columbia University, 200 Mudd, New York, NY 10027, USA.
(miw2103@columbia.edu)}

\begin{document}

%
%

\title{A Multiscale Model of Partial Melts 2: Numerical Results}
%

%
%


\author{G. Simpson}
\affil{Department of Mathematics, University of Toronto, Toronto, Ontario, Canada.}

\author{M. Spiegelman}
\affil{Department of Applied Physics and Applied Mathematics,
Columbia University, New York, New York, USA}

\author{M. I. Weinstein}
\affil{Department of Applied Physics and Applied Mathematics,
Columbia University, New York, New York, USA}



%
%
%

%
%


\begin{abstract}
  In a companion paper, equations for partially molten media were
  derived using two-scale homogenization theory.  This approach begins
  with a grain scale description and then coarsens it through multiple scale
  expansions into a macroscopic model.  One advantage of
  homogenization is that effective material properties, such as
  permeability and the shear and bulk viscosity of the two-phase
  medium, are characterized by \emph{cell problems}, boundary value 
  problems posed on a representative microstructural cell.  The solutions 
  of these problems can be averaged to obtain macroscopic parameters
  that are consistent with a given microstructure. This is
  particularly important for estimating the ``compaction length''
  which depends on the product of permeability and bulk viscosity and
  is the intrinsic length scale for viscously deformable two-phase flow.

  In this paper, we numerically solve ensembles of cell problems for
  several geometries. We begin with simple
  intersecting tubes as this is a one parameter family of problems
  with well known results for permeability.  Using this data, we
  estimate relationships between the porosity and all of the effective
  parameters with curve fitting.  For this problem, permeability
  scales as $\phi^n$, $n\sim2-3$, as expected and the bulk viscosity scales as
  $\phi^{-m}$, $m\sim 1$, which has been speculated, but never shown directly for
  deformable porous media. The second set of cell problems add
  spherical inclusions at the tube intersections.  These
  show that the permeability is controlled by the smallest pore
  throats and not by the total porosity, as expected.  The bulk
  viscosity remains inversely proportional to the porosity,
  and we conjecture that this quantity is insensitive to the specific
  microstructure.  The computational machinery developed
  can be applied to more general geometries, such as texturally
  equilibrated pore shapes.   However, we suspect that the
  qualitative behavior of our simplified models persists in these more 
  realistic structures.  In particular, our hybrid
  numerical--analytical model predicts that for purely mechanical
  coupling at the microscale, all homogenized models will have a
  compaction length that vanishes as porosity goes to zero.  This has
  implications for computational models, and it suggests that these
  models might not resist complete compaction.
\end{abstract}

%
%

%

\begin{article}

%
%

\section{Introduction}
Partially molten regions in the Earth's asthenosphere (e.g. beneath
mid-ocean ridges or subduction zones) are usually  modeled as a
viscously deformable permeable media.  Such models are typically
composed of macroscopic equations for the conservation of mass,
momentum and energy of each phase.  In our companion paper,
\cite{simpson08a}, \nocite{simpson08b} 
we derived several systems of governing equations
for partially molten systems  using homogenization.  Briefly, we began with a grain scale description of two interpenetrating fluids, each satisfying the Stokes equations, coupled by their common interface.  Several different macroscopic models could then be coarsened from this microscopic description, depending on our assumptions on the velocities, viscosities, and grain scale geometry.

An important feature of this approach is that macroscopic properties
such as permeability, shear viscosity, and bulk viscosity naturally
appear in the macroscopic equations, even if they are not defined at
the grain scale.  In particular, permeability and bulk viscosity are
properties of the two-phase aggregate, not the volume
averages of small scale variations. In contrast, previous work on the
magma problem, including \cite{mckenzie1984gac, bercovici2003etp,
  bercovici2005tpg, bercovici2007md, hiermajumder2006rgb,
  ricard2007pmc}, began with models much larger than the grain scale.
There, the viscosities, permeability, and other closures were assumed and
justified from other results. In contrast, homogenization derives these properties self-consistently.

While homogenization techniques appropriately inserts the
constitutive relationships into the macroscopic equations, connecting them 
to the microstructure requires the solution of
specific ``cell problems.'' For the physical system derived in
\cite{simpson08a}, there are actually a series of ten Stokes problems
for both fluid and solid posed on the micro-scale (which can be
reduced to four for micro-structures with sufficient symmetry).

In this paper, we numerically explore the cell problems to extract
parameterizations of the various effective parameters in terms of porosity, 
a simple measurement of the microstructure.
%
A feature of this work is to derive constitutive relationships for the bulk
viscosity as a function of porosity, which is essential for describing
compactible permeable media.  In particular, we show that for a range
of simple pore structure the effective
bulk viscosity is  related to the porosity as.
\[
\zeta_\eff \propto \f^{-1}
\]
Our results and some additional theory suggest that
this scaling is insensitive to the specific pore geometry.  

Since permeability and bulk-viscosity can both be consistently
related to the same microstructure, these calculations also allow us
to study the ``compaction length'' \cite[]{mckenzie1984gac}.  At sufficiently low melt concentrations, it is approximately
\begin{displaymath}
  \delta_\comp\propto\sqrt{k_\eff\zeta_\eff}  
\end{displaymath} 
which depends on the product of the derived permeability $k_\eff$ and
bulk viscosity $\zeta_\eff$, both of which depend on the
porosity. The compaction length is the intrinsic length
scale in magma dynamics.   This work suggests that under solely mechanical deformation,
\[
\lim_{\f\to 0} \delta_\comp(\f) = 0,
\]
implying that no mechanical mechanism prevents a region from
compacting to zero porosity. This result also places strong resolution
constraints on computational models of magma migration which may
require a regularization for small porosities. 

An outline of this work is as follows.  In Section \ref{sec:review}, we review 
several models of partially molten rock and highlight the constitutive relations.  
In Section \ref{sec:tubes}, we demonstrate the process of assuming a cell 
geometry to extract computational closures for the macroscopic system.  
We revisit these closures in Section \ref{sec:general_domains} for more 
general geometries to assess their robustness.  Finally, in Section 
\ref{sec:discuss}, we combine our numerics with the equations and 
examine the implications.

\section{Review of Equations and Constitutive Relations}
\label{sec:review}

\subsection{Macroscopic Equations}
In \cite{simpson08a}, we showcased three models for momentum
conservation in a partially molten medium.  They were distinguished by
the assumed scalings for the relative velocities and viscosities
between the fluid and solid phases, along with the connectivity of the
pore network.  One of them, dubbed Biphasic-I, was given by the
equations:
\begin{subequations}
\begin{gather}
\label{eq:bi1}
\begin{split}
0&= \overline{\rho}\bg-\nabla P +\nabla\bracket{\paren{\zeta_\eff-\frac{2}{3} \mu_s(1-\f)  } \nabla \cdot \bV^s }\\
&\quad+ \nabla \cdot\bracket{2{(1-\f)\mu_s }e(\bV^s)   }+ \nabla \cdot \bracket{2\eta_\eff^{lm} e_{lm}(\bV^s) }
\end{split}\\
\label{eq:bi2}
\f(\bV^f -\bV^s) = -\frac{k_\eff}{\mu_f}\paren{\nabla P - \bg^f}\\
\label{eq:bi3}
\nabla \cdot\bracket{\f \bV^f+(1-\f)\bV^s}=0
\end{gather}
\end{subequations}
Notation for his model may be found in Table \ref{table:effective}. In
particular, $k_\eff$ and $\zeta_\eff$ are the emergent permeability and bulk
viscosity.  $\eta_\eff$ is an auxiliary, tensorial, shear viscosity
capturing grain scale anisotropy.  All are related to the
aforementioned cell problems.    We highlight this case because (\ref{eq:bi1}
-- \ref{eq:bi3}) is nearly identical to the models of McKenzie,
Bercovici, Ricard, and others in the absence of melting and surface
physics.

 \begin{table}
  \centering
  \caption{Notation for macroscopic equations derived by homogenization.}
  \label{table:effective}
  \begin{tabular}{rp{14cm}}
  \hline\hline
  Symbol & Meaning \\
  \hline
  $\delta_\mathrm{comp.}$ & Compaction length\\
  $e(\bv)$ & Strain rate tensor, $e(\bv) = \frac{1}{2}(\nabla\bv + (\nabla\bv)^T)$\\
  $\eta_\eff$ & Supplementary anisotropic viscosity derived by homogenization\\
$\f$ & Porosity\\
$K$ & Permeability tensor of the matrix derived by homogenization\\
$k_\eff$ & Isotropic permeability of the matrix derived by homogenization\\
$\mu_f$ & Shear viscosity of the melt \\
$\mu_s$ & Shear viscosity of the matrix \\
$P$ & Macroscopic (fluid) pressure derived by homogenization\\
  $\rho_f$&Melt density \\
  $\rho_s$&Matrix density\\
  $\overline{\rho}$&Mean density, $\overline{\rho} = \rho_f\f + (1-\f)\rho_s$\\
$\mathbf{V}^f$ & Macroscopic fluid velocity derived by homogenization\\
$\mathbf{V}^s$ & Macroscopic solid velocity derived by homogenization\\
$\zeta_\eff$ &Bulk viscosity of the matrix derived by homogenization\\
    \hline
  \end{tabular}
\end{table}

\subsection{Constitutive Relations}
\label{sec:relations}
The constitutive relations for the permeability and viscosity are
fundamental to the dynamics of these models.  Indeed, they are the
source of much nonlinearity and it is useful to review some proposed 
closures.  Notation for this appears in Table \ref{table:constitutive}.

 \begin{table}
  \centering
  \caption{Notation for constitutive relations in other models.}
  \label{table:constitutive}
  \begin{tabular}{rp{14cm}}
  \hline\hline
  Symbol & Meaning \\
  \hline
  $\phi_\ast$ & Critical porosity for activation of $\mu_{s+f}$ viscosity\\
  $\kappa$ & Permeability of the matrix \\
 $\mu_{s+f}$ & Shear viscosity of the matrix in the presence of melt\\
 $\zeta_s$ & Bulk viscosity of the matrix\\
    \hline
  \end{tabular}
\end{table}

At low porosity, it is common to relate permeability, $\kappa$, to porosity by a power law, $\kappa \propto \phi^n$.  Estimates of $n$ vary, $n \sim 2 - 5$, \cite{scheidegger74pft, bear1988dfp, dullien1992pmf, turcotte2002g, mckenzie1984gac, doyen1988pca, cheadle1989pte, martys1994usf, faul1994ibm, faul1997ppm, faul2000cmd, koponen1997pae, wark1998gsp, wark2003rps, cheadle2004pta}.   For partially molten rocks, the exponent is better constrained by both analysis and experiment  to $n \sim 2 - 3$.  

For the matrix shear viscosity, \cite{hirth1995ecd1, hirth1995ecd2, kohlstedt2000rpm, kelemen1997rmm, kohlstedt2007prm} experimentally observed a melt weakening effect, which they fit to the curve:
\begin{equation}
\mu_{s+f} = \mu_s \exp\paren{-\f/\f_\ast}, \quad \f_\ast = O(10^{-2})
\end{equation}
$\mu_{s+f}$ is the shear viscosity of the solid matrix in the presence
of melt; $\mu_s$ is the shear viscosity a melt-free matrix.  In
\cite{bercovici2001a, bercovici2003etp} and related works, the
viscosity is weighted by $(1-\f)$, which is also present in
\eqref{eq:bi1}.  Reiterating, $\mu_{s+f}$ is a fitting of experimental
data.  Regardless, the shear viscosity is taken to be isotropic, and
porosity weakening in other models.

Lastly, the bulk viscosity, $\zeta_s$, is often taken as $\zeta_s \propto
\f^{-m}$, $m \sim 0-1$, though $m$ is usually either zero or one.
$m=0$ has often been used because the variation of bulk viscosity with
porosity is poorly constrained by observations.  \cite{scott1984ms}
and others invoked the bore hole studies of ice by \cite{nye1953flow} to justify $m=1$.  In
that work, Nye considered the dynamics of individual
spherical and cylindrical voids in an infinite medium.
\cite{taylor1954tcv, prudhomme78} computed $m=1$, in the limit of
small porosity, using models of incompressible fluids mixed with gas
bubbles.  \cite{schmeling2000pma} also employed $\zeta_s \propto \f^{-1}$,
citing studies for the analogous question of the effective bulk modulus of 
a fluid filled poroelastic medium.  Those
results rely on the self-consistent approximation methodology,
treated in \cite{torquato2002rhm}.  Models in
\cite{bercovici2001a, bercovici2003etp} and related works possess a
$\phi^{-1}$ term that functions as a bulk viscosity, though it has a
very different origin.  \cite{mckenzie1984gac} suggested using a
metallurgical model of spheres due to \cite{arzt1983pah} which
considered a range of mechanisms for densification of powders under
hot isostatic pressing.  For viscous compaction mechanisms, the Arzt
result recover a bulk viscosity proportional to $\f^{-1}$ for Newtonian
viscosities.  However, for very small porosities, they suggest that
surface diffusion effects become important and imply that $\zeta_s
\propto \log(\f^{-1})$.  Finally, \cite{connolly1998compaction} invoke
an assymetric bulk viscosity that is weaker during expansion than
during compaction, which is motivated by the deformation of brittle
crustal materials. However, it is unclear if this rheology is relevant to
high-temperature/high-pressure creeping materials as are expected in
the mantle. In general, it is expected that the
bulk-viscosity should become unbounded as the porosity reduces to zero
as the system then becomes incompressible.

\subsection{Cell Problems}
In homogenization, a medium with fine scale features is modeled by 
introducing two or more spatial scales.  As discussed in \cite{simpson08a}, 
the direct approach makes multiple scale expansions of the dependent 
variables, letting them depend on both the coarse and fine length scales.

The two characteristic lengths in our model are $L$, the macroscopic 
scale, and $\ell$, the grain scale.  Their ratio,
\[
\eps = \frac{\ell}{L}
\]
is the parameter in which we perform the multiple scale expansions.  
In the process of performing the expansions and matching powers of 
$\eps$, we are mathematically constrained to solve a collection of 
auxiliary cell problems posed on the fine scale.  Notation for this setup
is provided in Table \ref{table:cell_problem_structure}.

  \begin{table}
  \centering
  \caption{Notation for cell problems.}
  \label{table:cell_problem_structure}
  \begin{tabular}{rp{14cm}}
  \hline\hline
  Symbol & Meaning \\
  \hline
  $\eps$ & Ratio of microscopic and macroscopic length scales, $\eps = \ell/L$\\
  $e_{y}(\bv)$ & Strain rate tensor, $e_{y}(\bv) = \frac{1}{2}(\nabla_y \bv + (\nabla_y \bv)^T)$\\
  $\gamma$& Interface between melt and matrix within cell $Y$\\
  $\ell$& Grain length scale\\
  $L$ & Macroscopic length scale\\
  $\nabla_y$ & Gradient taken with respect to the $\by$ argument\\
  $\nabla_y \cdot $ & Divergence taken with respect to the $\by$ argument\\
  $\Omega$ & Macroscopic region containing both melt and matrix\\
  $\by$ & Coordinate within the cell, $Y$\\
  $Y$ & The unit cell\\
  $Y_f$ & Portion of unit cell occupied by melt\\
  $Y_s$ & Portion of unit cell occupied by matrix\\
  $\zeta$& Pressure of the cell problem for a unit forcing on the divergence equation\\
    \hline
  \end{tabular}
\end{table}

\begin{figure}
\noindent\includegraphics{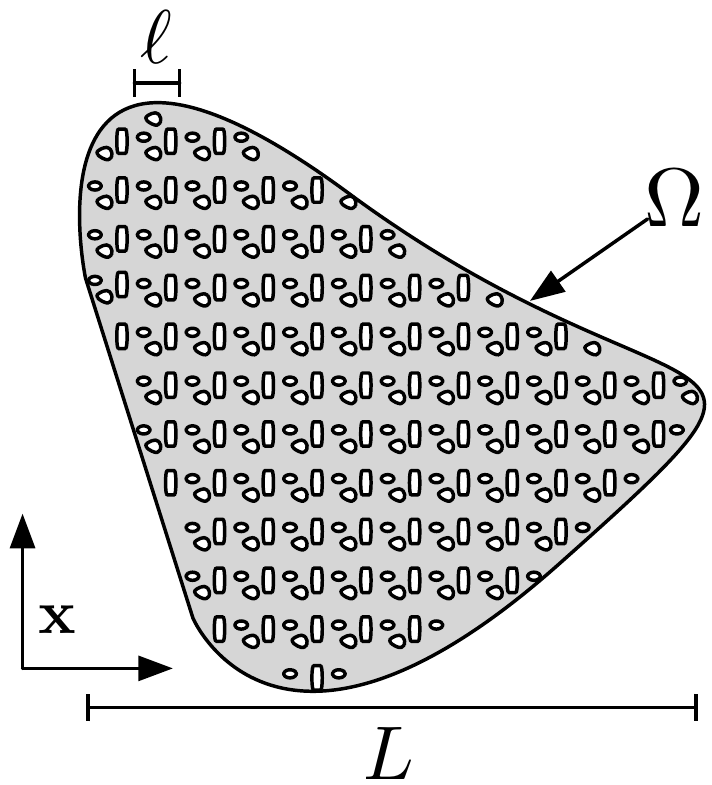}
\caption{The macroscopic domain $\Omega$.  The gray body is 
occupied by the matrix and the white inclusions are the melt.}
\label{fig:homog_domain}
\end{figure}

\begin{figure}
\noindent\includegraphics{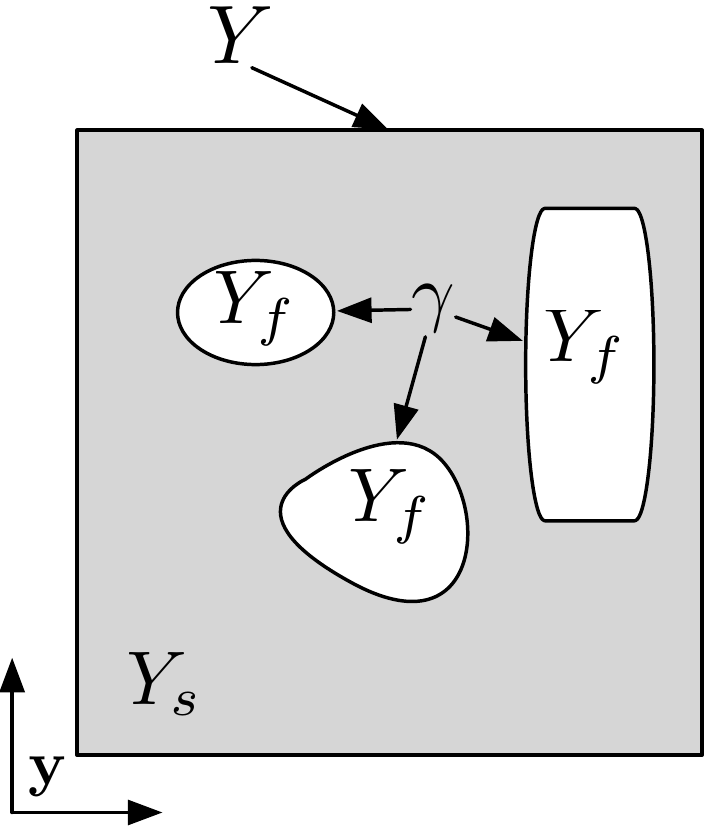}
\caption{The cell domain, $Y$, divided into fluid and solid regions, $Y_f$ 
and $Y_s$.  The two phases meet on interface $\gamma$.}
\label{fig:cell_domain}
\end{figure}

It is analytically advantageous to approximate the mixture as a periodic 
medium; such a configuration appears in Figure \ref{fig:homog_domain}.  
$\Omega$, the macroscopic region containing both matrix and melt is 
periodically tiled with scaled copies of the cell, scaled to unity in Figure 
\ref{fig:cell_domain}.  The cell problems are posed within  $Y_s$, the 
matrix portion of the cell, and $Y_f$, the melt portion for the cell.  
$Y_s$ and $Y_f$ meet on interface $\gamma$.

Generically, the cell problems take the form:
\begin{subequations}
\begin{align}
\label{eq:generic1}
\nabla_y\cdot\paren{-pI + 2 e_{y}(\mathbf{v})}&=\mathbf{f}\quad\text{in $Y_f$ or $Y_s$}\\
\label{eq:generic2}
\nabla_y \cdot \mathbf{v}&= g \quad \text{in $Y_f$ or $Y_s$}\\
\label{eq:generic3}
\paren{-pI+ 2 e_{y}(\mathbf{v})}\cdot \bn &=\tau \cdot \bn \quad\text{or}\quad\mathbf{u} =\mathbf{U} \quad \text{on $\gamma$}
\end{align}
\end{subequations}
which are generally, compressible Stokes flow problems for the microscopic solid
and melt velocities and pressures $(\mathbf{v}, p)$. Here,  $\nabla_y$ is the gradient, $\nabla_y \cdot$ is the divergence, and $e_{y}$ the strain-rate operator defined on the fine length scale $y$.  A more complete
 description is given in \cite{simpson08a}. $\mathbf{f}$, $g$, $\tau$,
and $\mathbf{U}$ are prescribed forcing functions on the relevant portion of
$Y$, either $Y_s$ or $Y_f$. The solution, $(\mathbf{v}, p)$,  is
periodic on the portion of the boundary not intersecting the interface
$\gamma$.  The cell problems may be interpreted as the fine scale
response to an applied stress on a unit cell
of either the melt or the matrix.  The material parameters --
$k_\eff$, $\zeta_\eff$, and $\eta_\eff$ -- are then defined as the
cell average of an appropriate manipulation of these pressures and
velocities defined at the pore scale.  In particular, an average of
the solid pressure in one problem determines the bulk viscosity and certain averages of
melt velocities in another problem determine the permeability.  Again, we emphasize
that the material parameters are not volume averages of the same
parameters at the fine scale.

  \begin{table}
  \centering
  \caption{Notation for cell problems, continued.}
  \label{table:cell_problems_continued}
  \begin{tabular}{rp{14cm}}
  \hline\hline
  Symbol & Meaning \\
  \hline
  $\bar{\chi}^{lm}$ & Velocity of the cell problem for a unit shear stress forcing on the solid in the $lm$ component of the stress tensor\\
  $\mathbf{e}_i$ & Unit vector in the i--th coordinate, $\mathbf{e}_1^T = (1, 0, 0)$\\
  $\f_\eff$ & Effective porosity, portion of the porosity in which there is appreciable flow.  $\f_\eff \leq \f$. \\
  $\mathbf{k}^i$& Velocity of the cell problem for a unit forcing on the fluid in the   $\mathbf{e}_i$  direction\\
  $\mathbf{k}_1^1$& First component of the velocity from the cell problem with unit forcing on the fluid in the $\mathbf{e}_1$ direction\\
  $\bar{\xi}$ & Velocity of the cell problem for a unit forcing on the divergence equation\\
  $\mean{\cdot}_f$ & Volume average of a quantity over the melt portion of a cell, $\mean{\cdot}_f = \int_{Y_f} \cdot d \by$\\
  $\mean{\cdot}_s$ & Volume average of a quantity over the matrix portion of a cell, $\mean{\cdot}_s = \int_{Y_s} \cdot d \by$\\
  $\pi^{lm}$& Pressure of the cell problem for a unit shear stress forcing on the solid in the $lm$ component of the the stress tensor\\
  $ q^i$&Pressure of the cell problem for a unit forcing on the fluid in the   $\mathbf{e}_i$ direction\\
  $\zeta$& Pressure of the cell problem for a unit forcing on the divergence equation\\
    \hline
  \end{tabular}
\end{table}


\section{Effective Parameters: Intersecting Tube Geometry}
\label{sec:tubes}
To connect the effective material properties 
-- $\eta_\eff$, $k_\eff$, and $\zeta_\eff$ -- with the porosity, we must solve 
the cell problems and extract a parameterization.  Unfortunately, the 
solution of the Stokes equations in a generic  three dimensional domain 
lacks an analytic representation.  Thus we compute the
solutions to the cell problems numerically and fit the results to
appropriate parameterized constitutive models.  Notation for this
section is summarized in Table \ref{table:cell_problems_continued}.

\begin{figure}
\noindent\includegraphics{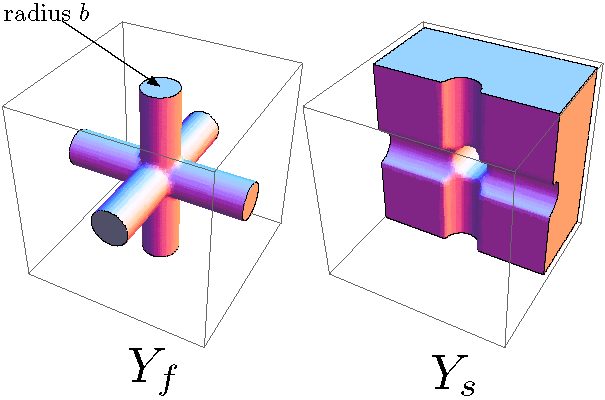}
\caption{A cell geometry composed of triply intersecting cylinders of equal radius.}
\label{fig:tubes}
\end{figure}

As a first example of numerically closing the constitutive relations in a 
homogenization based model, we study the cell domain of triply intersecting
cylinders, pictured in Figure \ref{fig:tubes}.  The fluid occupies the
cylinders while the solid is the complementary portion of the cube.
While this geometry is an oversimplification of real pore-geometries,
it serves as proof of concept of the unified, self-consistent homogenization
algorithm.  Additionally, it serves as a numerical check since the permeability for
this problem is expected to scale as $\phi^2$.  In Section \ref{sec:general_domains}
we examine a  generalization of this geometry.

In what follows, we emphasize that the information given by the parameterizations
should be read with a certain skepticism.  The order of magnitude and signs
of the coefficients and exponents are of greater utility than the particular numbers.

We remark here that for the intersecting tube geometry, the tube
radius scaled to the cell-size, $b$, can be explicitly related to the
porosity:
\begin{equation}
\label{eq:tube-porosity}
\f = 3 \pi b^2 -8\sqrt{2} b^3
\end{equation}

We solve our problems using finite elements on unstructured meshes
using the open source libraries from the FEniCS and PETSc projects
\cite[e.g.][]{DupHof2003,KirLog2007,Log2007,petsc-user-ref,petsc-web-page},
with meshes generated using CUBIT \cite[]{cubit}.  Details of this method
and numerical benchmarks are given in Appendix
\ref{sec:computation_notes}.

\subsection{Effective Permeability}
\label{sec:perm}
The first cell problem we treat is for permeability.  The equations are:
\begin{subequations}
\begin{align}
-\nabla_y q^i + \nabla_y^2 \mathbf{k}^i &= -\mathbf{e}_i\quad\text{in $Y_f$}\\
\nabla_y \cdot \mathbf{k}^i &= 0\quad\text{in $Y_f$}\\
\mathbf{k}^i &= 0\quad \text{on $\gamma$}
\end{align}
\end{subequations}
where $\mathbf{k}^i$ is a three-dimensional velocity and $q^i$ is a scalar pressure, \emph{for each $i=1, 2, 3$}.
These are equations (64a -- 64c) in \cite{simpson08a} and describe the motion of the melt through the 
porous matrix. 
In general, the permeability is the second order tensor:
\[
\mean{K}_f= \begin{bmatrix} \int_{Y_f}{\mathbf{k}^1}d\by & \int_{Y_f}{\mathbf{k}^2}d\by & \int_{Y_f}{\mathbf{k}^3}d\by\end{bmatrix}
\]
The symmetry properties of the domain simplify this to:
\[
\mean{K}_f = \mean{\bk_1^1}I,\quad\textrm{$\bk_1^1$ is the first component of vector $\mathbf{k}^1$}
\]
Thus, it is sufficient to compute the case $i=1$.  Then $k_\eff$, the permeability of the matrix, in \eqref{eq:bi2} is
\begin{equation}
\boxed{k_\eff \equiv \mean{\bk_1^1}_f}
\end{equation}

Darcy's Law and permeability have been studied by many techniques, including homogenization; we refer the reader to the references in Section \ref{sec:relations}.  We study it here to understand how the permeability behaves \emph{in concert} with the other constitutive relations as the microstructure varies.  This also serves as a benchmark problem for our software; see Appendix \ref{sec:computation_notes}.  

As noted, porosity and permeability are often related by a power law, $\kappa \propto \f^n$, with $n \sim 2 - 5$.  To motivate such a relation, we turn to a toy model, as presented in \cite{turcotte2002g}.  The melt is assumed to be in Poiseuille flow through triply intersecting cylinders.  Additionally, the cylinders have small radii; it is a low porosity model.  The permeability of such a system is
\begin{equation}
\label{eq:toy_perm}
\kappa_{\textrm{toy-I}} = \frac{\ell^2 \f^2}{72 \pi}\approx 0.0044 \ell^2 \f^2.
\end{equation}
Other simple models are developed in \cite{scheidegger74pft, bear1988dfp, dullien1992pmf}.  

\begin{figure*}[t]
\begin{center}
\includegraphics{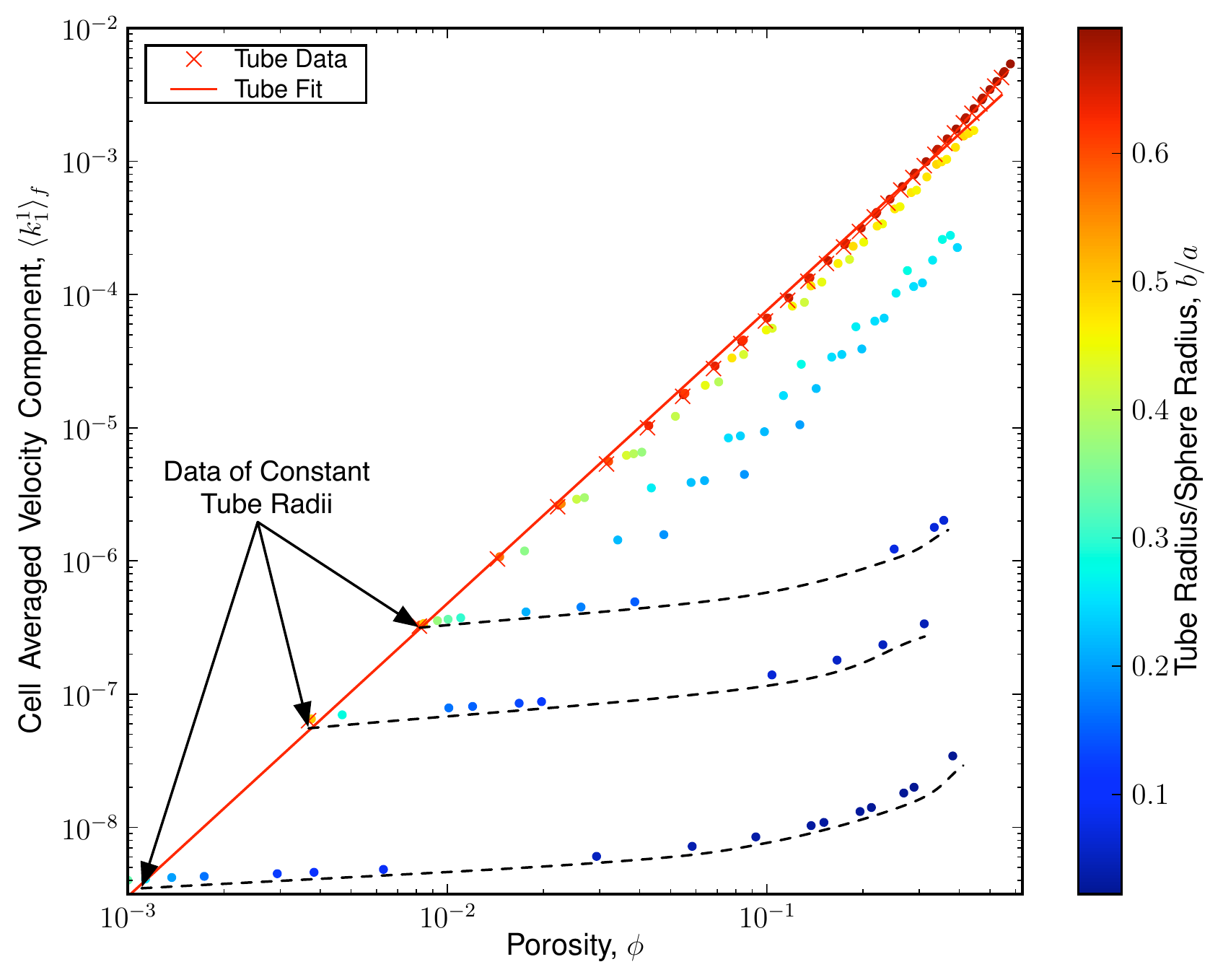}
\caption{Numerically computed values of $\mean{\bk_1^1}_f$, the effective permeability, plotted against porosity for both the tube geometry and the sphere+tube geometry.  The scattered circles are data from sphere+tube geometries, colored by the ratio of tube radius to sphere radius. The tube geometry offers an upper bound for a given porosity.}
\label{fig:perm-comparison}
\end{center}
\end{figure*}

We now  fit our computed permeabilities, $\mean{\bk_{1}^1}$,  to porosity by such a relation. For the tube domains, the least squares fit is
\begin{equation}
\label{eq:tube_porosity}
\mean{\bk^{1}_1}_f = \exp(-4.42 \pm .105)  \f^{2.20 \pm .0391}.
\end{equation}
This curve and the data appear in Figure \ref{fig:perm-comparison}.
The fit matches expectations of an $O(10^{-3}-10^{-2})$ prefactor and
an exponent $\sim 2- 3$.  The error in \eqref{eq:tube_porosity} is the
associated 95\% confidence interval.  We report these intervals in all
regressions, though they rely on the specious assumption that error in
our synthetic data is normally distributed.


\subsection{Effective Bulk Viscosity}
\label{sec:bulk}
The effective bulk viscosity is related to the solution $(\bar{\xi}, \zeta)$ of:
\begin{subequations}
\begin{align}
\label{eq:dilation1}
\nabla_{y}\cdot\paren{-\zeta I + 2 e_{y}(\bar{\xi})}&=0\quad\text{in $Y_s$}\\
\label{eq:dilation2}
\nabla_y \cdot \bar{\xi}&= 1\quad \text{in $Y_s$}\\
\label{eq:dilation3}
\paren{-\zeta I + 2 e_{y}(\bar{\xi})}\cdot \bn& =0\quad \text{on $\gamma$}
\end{align}
\end{subequations}
where $\bar{\xi}$ is a three-dimensional velocity and $\zeta$ is a scalar pressure. These are equations (B1a -- B1c) in \cite{simpson08a} and are associated with the compaction of the matrix. The effective bulk viscosity is then defined as
\begin{equation}
\label{eq:bulk_eff}
\boxed{\zeta_\eff \equiv\mu_s \mean{\zeta}_s-\frac{2}{3}\mu_s (1-\f)}
\end{equation} 

The dependence of the effective bulk viscosity of partially molten rock as a function of porosity is the most poorly constrained of the material properties.  This is partly due to the difficulties in constructing an experiment that will measure it as a function of porosity independently of the shear viscosity \cite{mckenzie1984gac, kelemen1997rmm, stevenson1991mfr}.  As mentioned in Section \ref{sec:relations}, a bulk viscosity $\propto\f^{-1}$ has often appeared in the literature.

Two toy models for the bulk viscosity of an incompressible fluid seeded with compressible gas bubbles were formulated by \cite{taylor1954tcv, prudhomme78}.  They relate the bulk viscosity to the porosity as:
\begin{subequations}
\begin{align}
\label{eq:bulkviscosity_taylor}
\zeta_s &=\frac{4}{3} \frac{\mu_s}{\phi}&\quad \textrm{Taylor}\\
\label{eq:bulkviscosity_prudhomme}
\zeta_s &=\frac{4}{3} \frac{\mu_s}{\phi}\paren{1-\f}&\quad \textrm{Prud'homme and Bird}
\end{align}
\end{subequations}
Taylor's expression, \eqref{eq:bulkviscosity_taylor}, relied on a
single inclusion model for a gas bubble in an infinite medium.
\eqref{eq:bulkviscosity_prudhomme} is derived by considering a sphere
of fluid with a gas filled spherical cavity, and seeking the bulk
viscosity of a compressible fluid that will give rise to the same
radial stress for specified boundary motion.  The $1-\f$ is due to
Prud'homme and Bird restricting their model to a finite volume.  This factor
also appears in the proposed bulk viscosity of
\cite{schmeling2000pma}.  
These expressions have motivated the use of
$\zeta_s \propto \f^{-1}$ in macroscopic models of partial melts,
although they actually arise from a different problem of a
compressible inclusion in an incompressible fluid, rather than the
divergent flow of two interconnected incompressible fluids.
Surprisingly, when we consider a simple toy problem to approximate the
the full cell calculation, we find that the expressions are identical.

\begin{figure*}[t]
\noindent\includegraphics{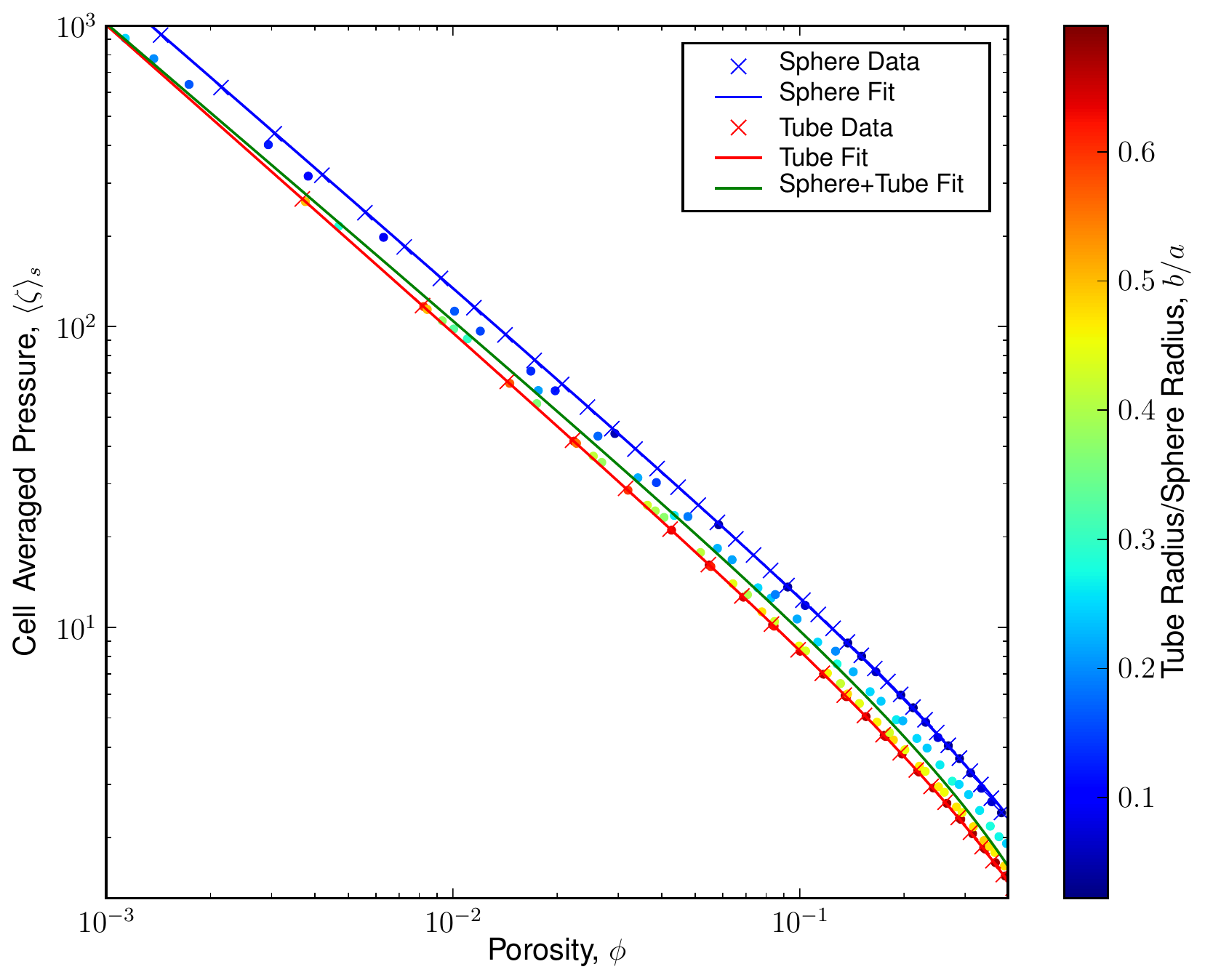}
\caption{Numerically computed data for the dilation stress cell problem, (\ref{eq:dilation1}--\ref{eq:dilation3}), on the three geometries, along with the least square fits \eqref{eq:bulk_tube_fit}, \eqref{eq:bulk_sphere_fit}, and \eqref{eq:bulk_st_fit}).  The scattered circles are data from sphere+tube geometries, colored by the ratio of tube radius to sphere radius.  The sphere+tube data is bounded between the the tube data and the sphere data.}
\label{fig:bulkcomparison_plot}
\end{figure*}

Appendix  \ref{sec:bulk_sphere} provides details of this toy problem,
which is  related to equations (\ref{eq:dilation1} --
\ref{eq:dilation3}), and gives the solution
\[
\mean{\zeta}_s = \frac{4}{ 3 \phi}\paren{ 1 + \frac{\f}{2}}\paren{1-\f},
\]
from which we get
\[
\zeta_\eff = \frac{4 \mu_s}{3 \f}\paren{1 -\f}.
\]
This  motivates trying to numerically fit $\mean{\zeta}_s$ to $(1-\f)^p/\f^q$, expecting $p$ and $q$ to be close to unity.  Indeed, the data, plotted in Figure \ref{fig:bulkcomparison_plot}, fits the curve
\begin{equation}
\label{eq:bulk_tube_fit}
\mean{\zeta}_s = \exp(-0.131 \pm 0.00514) \f^{-1.02 \pm 0.00132}(1-\f)^{0.884 \pm 0.00869}
\end{equation}
which is quite similar to the scaling of the toy model.

\subsection{Supplementary Anisotropic Viscosity}
\label{sec:supplementary_viscosity}
We now examine the cell problem related to the supplementary viscosity $\eta_\eff$, a fourth order tensor.  The equations are:
\begin{subequations}
\begin{align}
\nabla_y\cdot\paren{-\pi^{lm}I + 2 e_{y}(\bar{\chi}^{lm})}&=0\quad\text{in $Y_s$}\\
\nabla_y \cdot \bar{\chi}^{lm}&= 0\quad \text{in $Y_s$}\\
\paren{-\pi^{lm}\delta_{ij}+ 2 e_{y,ij}(\bar{\chi}^{lm}) } n_j &=-\frac{1}{2}\paren{\delta_{il}\delta_{jm}+\delta_{im}\delta_{jl}} n_j\quad \text{on $\gamma$}
\end{align}
\end{subequations}
where $\bar{\chi}^{lm}$ is a three-dimensional velocity vector and $\pi^{lm}$ is a scalar pressure, \emph{for each pair $(l,m)$, $l=1,2,3$ and $m=1,2,3$}.
These are equations (B2a -- B2c) from  \cite{simpson08a} and are tied to
tensorial surface stresses applied on the matrix. Using the solutions $(\bar{\chi}^{lm}, \pi^{lm})$,
\begin{equation}
\label{eq:supp_visc}
\boxed{\eta_\eff^{lm} \equiv \mean{e_y(\bar{\chi}^{lm})}_s}
\end{equation}
Because of symmetry, we need only consider two problems: $(l,m)=(1,1)$
and $(l,m)=(1,2)$, corresponding to normal stress and shear stress in
each direction.

\begin{figure*}[t]
\noindent\includegraphics{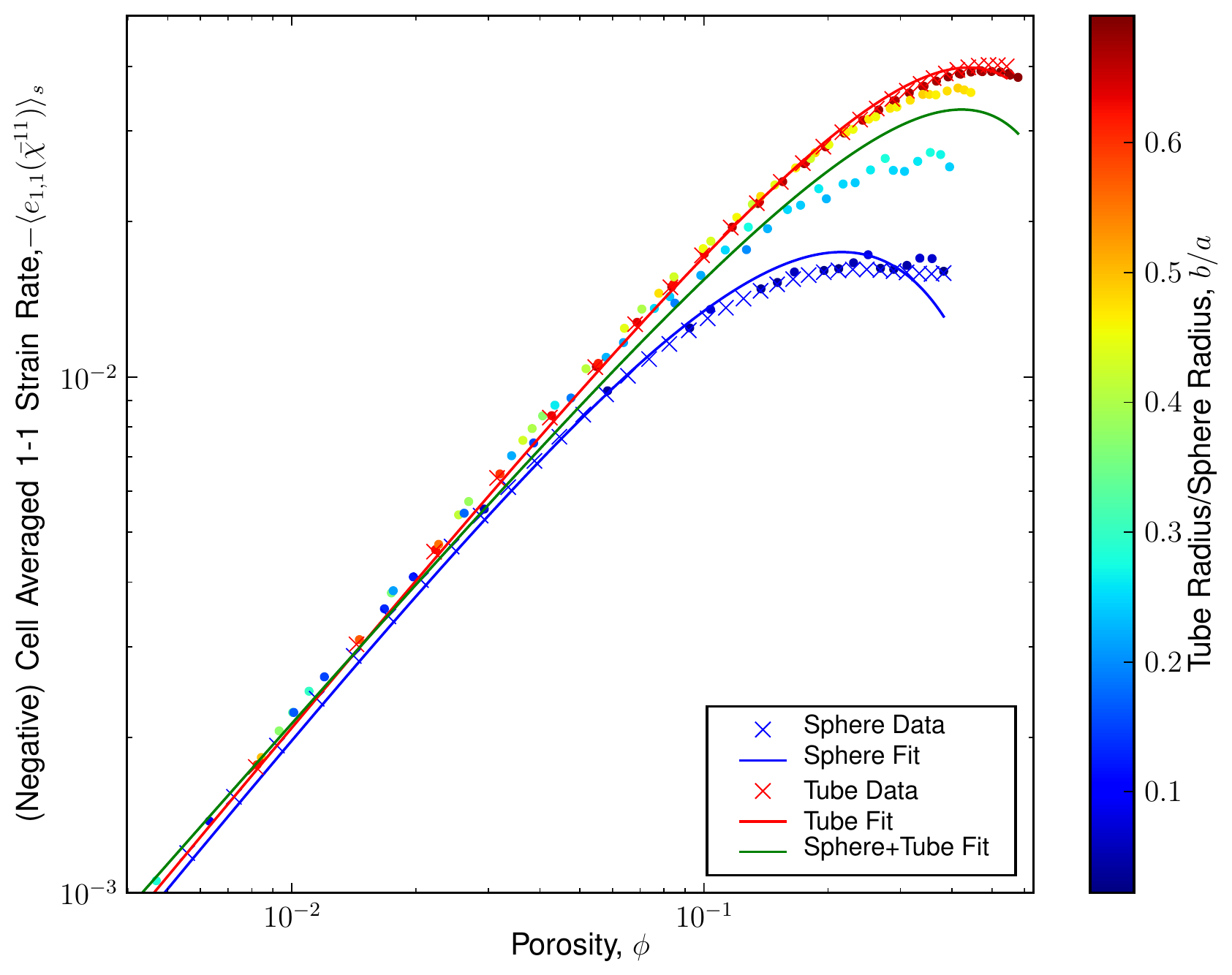}
\caption{Numerically computed data for the normal stress cell problem on the three geometries, along with the least square fits \eqref{eq:e11_tube},  \eqref{eq:e11_sphere}, and \eqref{eq:e11_st}.  The scattered circles are data from sphere+tube geometries, colored by the ratio of tube radius to sphere radius.  At small porosity there is little variation amongst the simulated domains.}
\label{fig:e11comparison}
\end{figure*}

Although we have no toy problem as motivation, $\f^p(1-\f)^q$ proved to be satisfactory.  First, we study the problem $(l,m)=(1,1)$, a uniaxial stress problem.  For the tubes, we fit
\begin{equation}
\label{eq:e11_tube}
-\mean{e_{1,1}(\bar{\chi}^{11})}_s = \exp(-1.72 \pm .0405) \f^{.964 \pm .0104}(1-\f)^{1.23 \pm .0685}
\end{equation}
This vanishes as $\f \to 0$ and as $\f \to 1$ and it is nearly linear at small porosity.  The curves and the data are plotted in Figure \ref{fig:e11comparison}.

\begin{figure*}[t]
\noindent\includegraphics{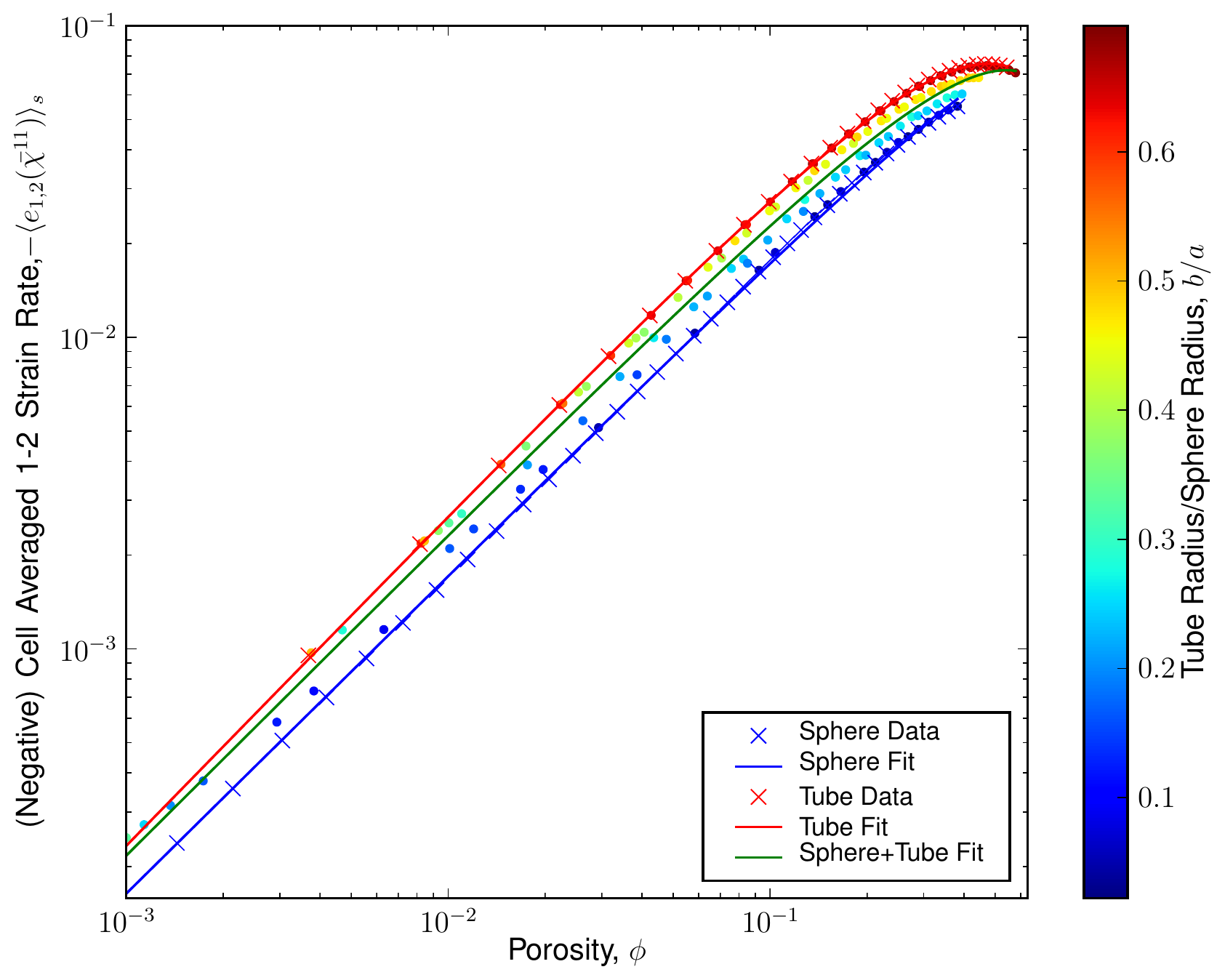}
\caption{Numerically computed data for the shear stress cell problem on the three geometries, along with the least square fits \eqref{eq:e12_tube}, \eqref{eq:e12_sphere}, \eqref{eq:e12_st}.  The scattered circles are data from sphere+tube geometries, colored by the ratio of tube radius to sphere radius.  Sphere+Tube data points are constrained between the sphere data and the tube data.  At all simulated porosities, there is less than an order of magnitude of variation.}
\label{fig:s12comparison}
\end{figure*}

There is also  the simple shear stress problem, $(l,m)=(1,2)$.  For this, we fit
\begin{equation}
\label{eq:e12_tube}
-\mean{e_{12}(\bar{\chi}^{(12)})}_s = \exp(-1.04 \pm .0188)\f^{1.06 \pm .00485}(1-\f)^{1.17 \pm .0318}
\end{equation}
Plots for this are given in Figure \ref{fig:s12comparison}.

We could now employ the simple computational closures \eqref{eq:tube_porosity}, \eqref{eq:bulk_tube_fit}, \eqref{eq:e11_tube}, and \eqref{eq:e12_tube}, in (\ref{eq:bi1} -- \ref{eq:bi3}), to simulate and study the macroscopic problem.

\section{Generalization of the Cell Domains}
\label{sec:general_domains}
\begin{figure}
\noindent\includegraphics[width= 20pc]{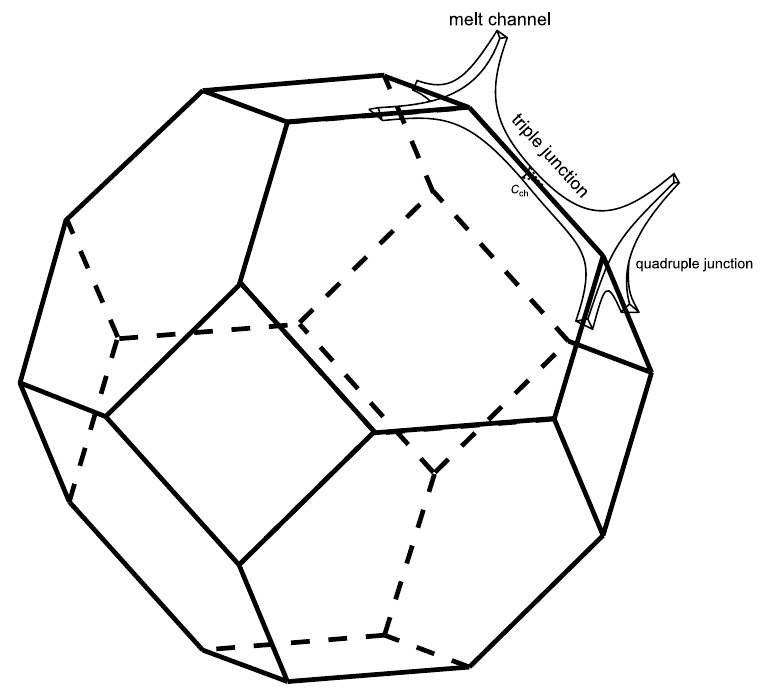}
\caption{An idealized Olivine grain from Figure 1 of \cite{zhu2003nmp}.  Melt channels are found at triple junctions, while melt pockets are found quadruple junctions.}
\label{fig:tetrakai}
\end{figure}

Regrettably, Earth materials are not as trivial as intersecting
cylinders.  Even an idealized olivine grain is a tetrakaidekahedron,
pictured in Figures \ref{fig:tetrakai}.  As depicted, some fraction of
the melt lies along the triple junctions and some is at the quadruple
junctions.  Other examples of idealized, texturally equilibrated
arrangements appear in \cite{vonbargen1986pia} and  \cite{cheadle1989pte, cheadle2004pta}.  These
methods are also amenable to studying \emph{random} media.  One could
compute relations based on ensembles of randomly generated cell
domains, solving all relevant cell problems on the ensemble.

\begin{figure}
\noindent\includegraphics{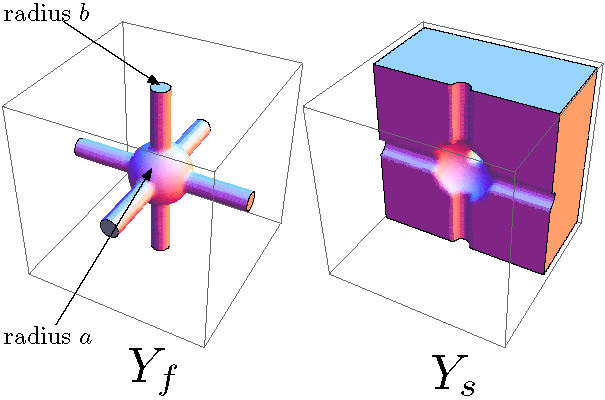}
\caption{A cell geometry composed of triply intersecting cylinders of equal radius, with a sphere at the intersection.}
\label{fig:sphere_tube_geometry}
\end{figure}

Motivated by Figure \ref{fig:tetrakai}, we explore a simple
generalization of the tube geometry by  adding a sphere of independent
radius at the intersection, as in Figure
\ref{fig:sphere_tube_geometry}.  This retains the symmetry of the
previous model, but adds a second parameter, allowing multiple
geometries for the same porosity.  The sphere captures some
aspect of the pocket at the quadruple junctions.  The sphere radius,
$a$, and the tube radius, $b$, are related to the porosity by the
equation:
\begin{equation}
\label{eq:tube-sphere-porosity}
\f =\pi\bracket{ -4 a^3 + 4a^2 \sqrt{a^2-b^2} + b^2 \paren{3 - 4 \sqrt{a^2-b^2}}+\frac{4}{3}a^3}
\end{equation}
We shall refer to it as the \emph{sphere+tube} geometry.  

We now repeat the computations of Section \ref{sec:tubes} on the sphere+tube geometry to assess the sensitivity of the effective parameters to cell geometry.   We also perform computations on domains where the fluid occupies an isolated sphere at the center of a cube.  Though this is disconnected, it provides useful information.

\subsection{Permeability Revisited}
\label{sec:perm_revisited}
Parameterizing the porosity-permeability relation for this generalization is quite challenging.  In contrast to the tube geometry, the data points of the sphere+tube geometry, plotted in Figure \ref{fig:perm-comparison}, do not collapse onto a curve.   There is some  positive correlation between permeability and porosity, and for a given porosity, {the permeability of the equivalent tube geometry is an upper bound}.

\begin{figure}
\begin{center}
\includegraphics[width=3in]{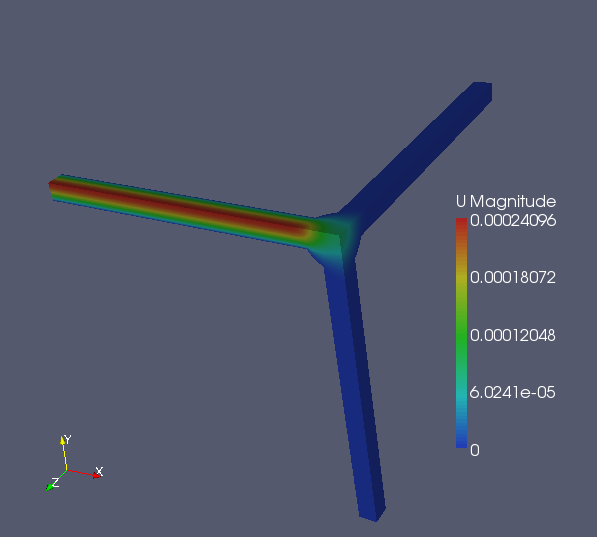}
\caption{The magnitude of the velocity for a permeability cell problem. This corresponds to the sphere+tube domain with sphere radius $a=.06$ and tube radius $b=.03$.}
\label{fig:flowfield1}
\end{center}
\end{figure}

\begin{figure}
\begin{center}
\includegraphics[width=3in]{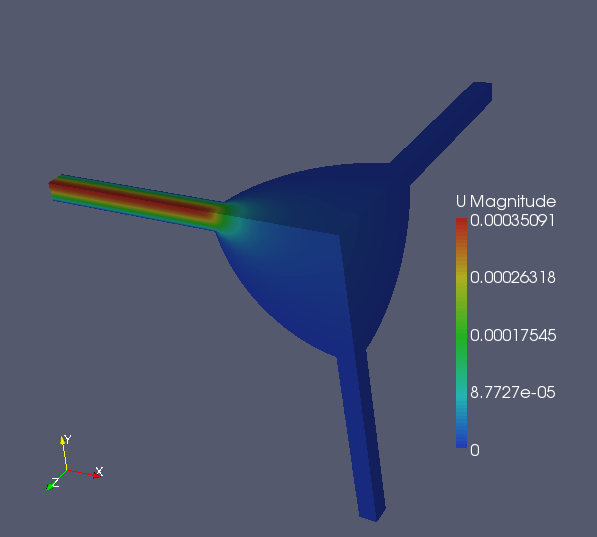}
\caption{The magnitude of the velocity for a permeability cell problem. This corresponds to the sphere+tube domain with sphere radius $a=.20$ and tube radius $b=.03$.}
\label{fig:flowfield2}
\end{center}
\end{figure}

To better understand the trend, we examine the computed flow fields in
Figures \ref{fig:flowfield1} and \ref{fig:flowfield2}.  These plot the
velocity magnitude on two fluid domains with the same tube size, but
different sphere sizes.  Most of the flow is within the tube.  While
there is some detrainment as it enters the sphere, the flow in the
tube appears insensitive to the size of the sphere.

\begin{figure*}[t]
\begin{center}
\includegraphics{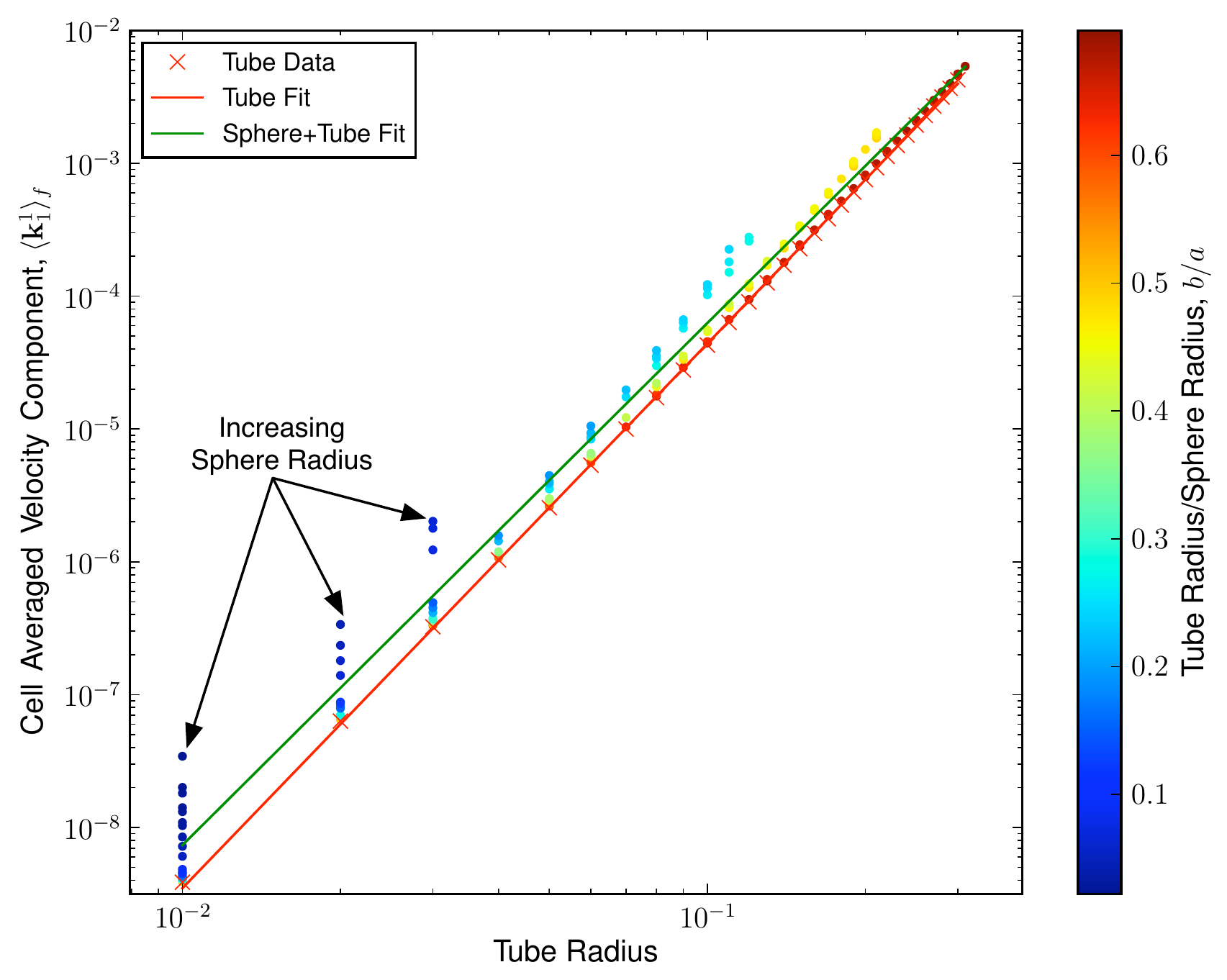}
\caption{$\mean{\mathbf{k}_1^1}_f$, the effective permeability, plotted against tube radius for both the tube geometry and the sphere+tube geometry.  The scattered circles are data from sphere+tube geometries, colored by the ratio of tube radius to sphere radius. The tube geometry offers a lower bound for a given porosity.}
\label{fig:darcy_tube_fit}.
\end{center}
\end{figure*}

This motivates  fitting permeability against tube radius.  Indeed, an alternative to \eqref{eq:toy_perm}, is
\begin{equation}
\label{eq:toy_perm_tube}
\kappa_{\textrm{toy-II}} = \frac{ \delta^4}{128 \ell^2}
\end{equation}
The tube diameter $\delta$, is equivalent to $2b$, $b$ the the tube radius in the tube and sphere+tube geometries.  Both data sets appear  in Figure \ref{fig:darcy_tube_fit}.  This is a significant improvement over Figure \ref{fig:perm-comparison}.  The least square fits are:
\begin{align}
\mean{\bk_1^1}_f &= \exp(-0.592 \pm .0354) b^{4.10 \pm .0156},\quad\text{for tube geometry}\\
\mean{\bk_1^1}_f &= \exp(-0.628 \pm .198) b^{3.93 \pm .0684},\quad\text{for sphere+tube geometry}
\end{align}
These estimates with \eqref{eq:toy_perm_tube}; taking $\delta=2b$ and scaling out $\ell$, this relationship is $k_{\textrm{toy-II}} = .125 b^4$.  The data is still positively correlated with sphere radius, altering it by as as much as an order of magnitude.  The deviations are greatest when both $b\ll1$ and $b \ll a$.

That the permeability is more strongly correlated with the tube radius than the overall geometry is not surprising.
\cite{koponen1997pae} discuss the notion of {effective porosity}, the portion of the void space where there is significant flow.  Denoting our effective porosity $\f_\eff$, we seek a relation $k_\eff \propto \f_\eff^n$. 

\begin{figure}
\begin{center}
\includegraphics[width=2in]{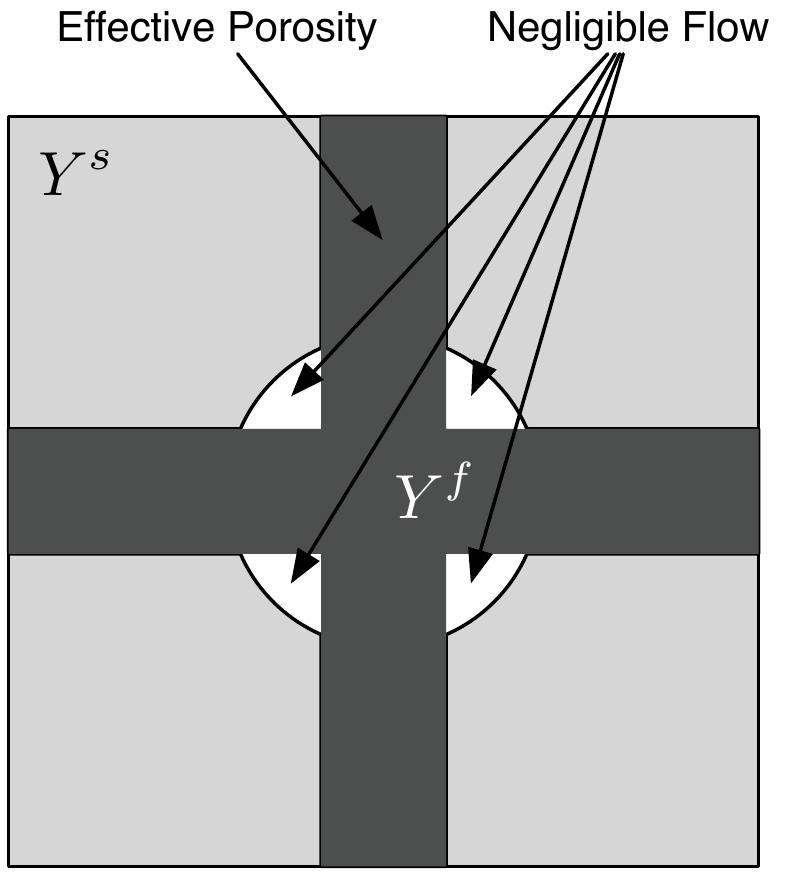}
\caption{The dark portion is the postulated effective porosity for the sphere+tube domains.}
\label{fig:effective_porosity}
\end{center}
\end{figure}

Given the flow fields in Figures \ref{fig:flowfield1} and \ref{fig:flowfield2} and the success with the tube radius fittings, we posit that the effective porosity is the portion of the porosity within the tubes.  A two-dimensional analog appears in Figure \ref{fig:effective_porosity}.   Using \eqref{eq:tube_porosity}, we define $\phi_\eff$ for the sphere+tube domains:
\begin{equation}
\label{eq:eff_porosity}
\f_{\mathrm{eff.}}=3 \pi b^2 - 8 \sqrt{2} b^3
\end{equation}
\cite{zhu2003nmp} made a similar approximation; from \cite{vonbargen1986pia} they construed that the permeability was controlled by the minimal cross-sectional area of the pore network.  A similar argument is made by \cite{cheadle1989pte}.  In our domains, the minimal cross-sectional area is $\pi b^2$.

\begin{figure*}[t]
\begin{center}
\includegraphics{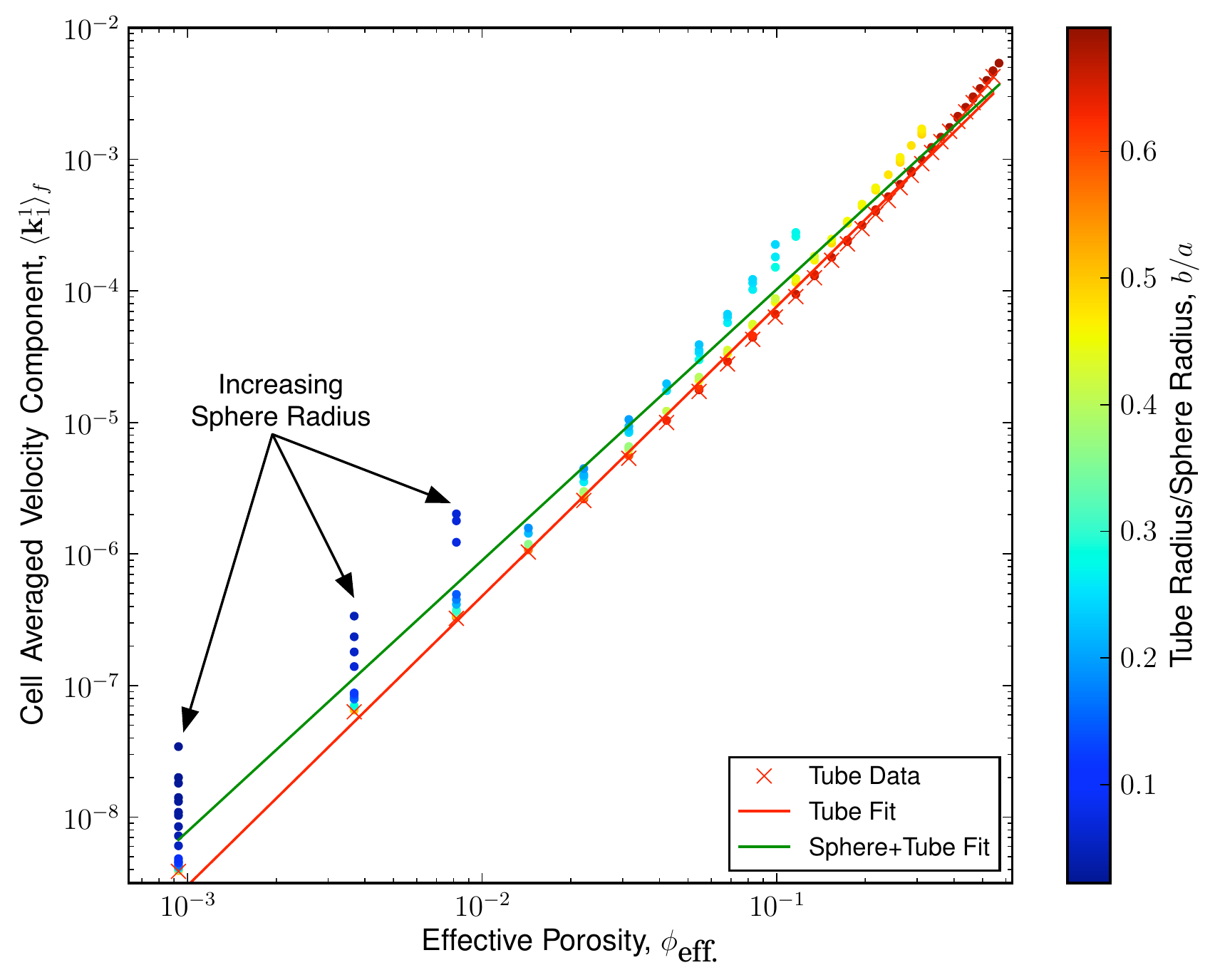}
\caption{$\mean{\mathbf{k}_1^1}_f$, the permeability, plotted against the effective porosity for the sphere+tube geometry.  The scattered circles are data from sphere+tube geometries, colored by the ratio of tube radius to sphere radius.  The tube data plotted against porosity also appears.}
\label{fig:effecitve_permeability}
\end{center}
\end{figure*}

We fit
\begin{equation}
\label{eq:phie_fit}
\mean{\bk_1^1}_f = \exp(-4.44 \pm .144) \f_{\eff}^{2.06\pm .0374}\quad\text{for sphere+tube}.
\end{equation}
This appears in Figure \ref{fig:effecitve_permeability}.  Again, deviation is highest for very large spheres with very thin tubes.  Unfortunately, $\phi_\eff$ does not satisfy a conservation law, making it a less than ideal macroscopic quantity to track, though it does satisfy the bound $\f_\eff \leq \f$.

\begin{figure*}[t]
\begin{center}
\includegraphics{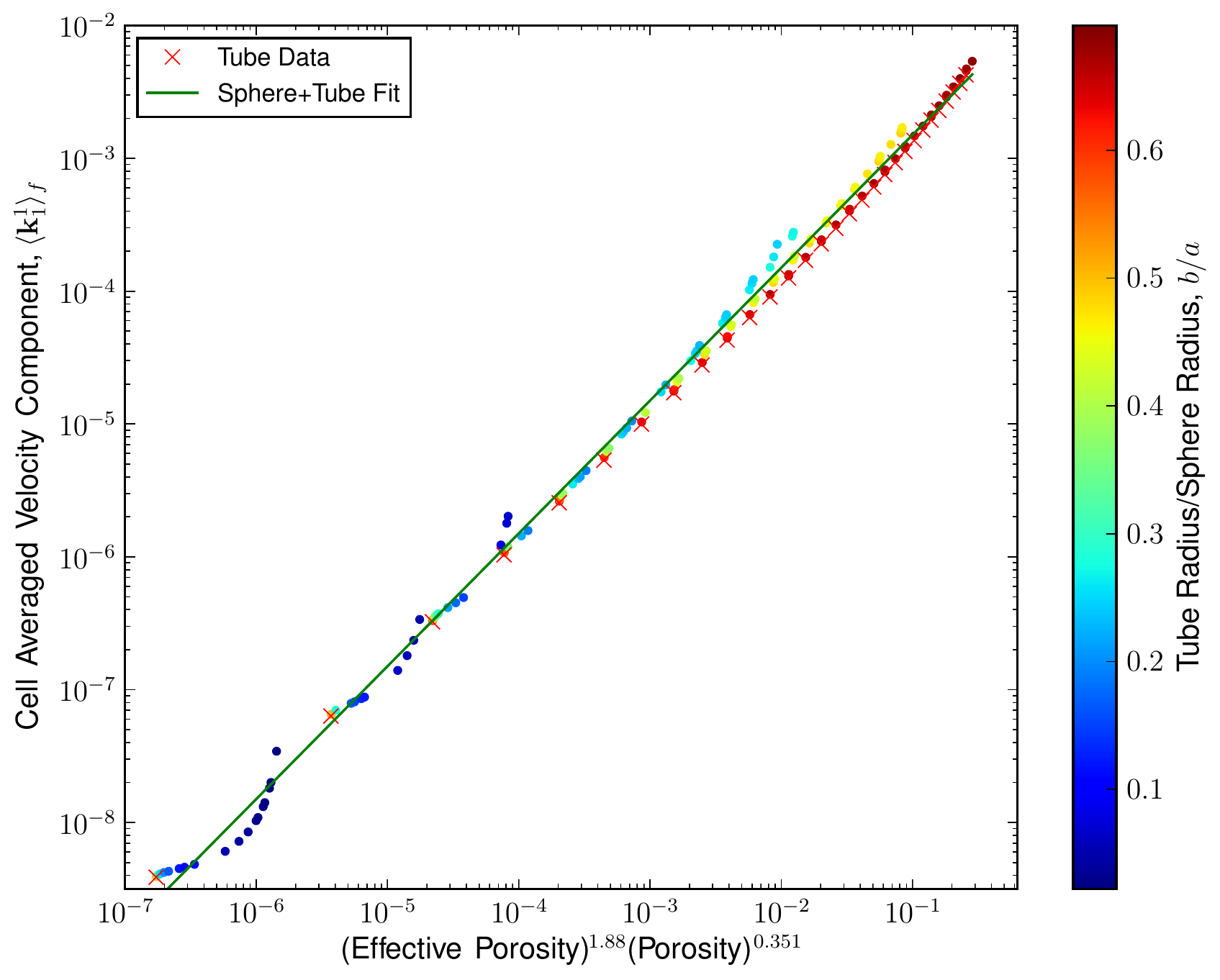}
\caption{$\mean{\mathbf{k}_1^1}_f$, the permeability, plotted against the effective porosity and porosity, using \eqref{eq:hybrid_fit}. The scattered circles are data from sphere+tube geometries, colored by the ratio of tube radius to sphere radius.  The tube data is also plotted, substituting $\phi$ for $\phi_\eff$ in \eqref{eq:hybrid_fit}.}
\label{fig:hybrid_fit}
\end{center}
\end{figure*}

There is still as much as an order of magnitude deviation at low porosity from relation  \eqref{eq:phie_fit}.  The unresolved part of the permeability for the $\phi_\eff$ fit is increasing in the sphere radius.  This motivates trying to fit against both $\f_\eff$ and another parameter.  It is sufficient to fit permeability to $\f_\eff$ \emph{and} $\f$, resulting in
\begin{equation}
\label{eq:hybrid_fit}
\mean{\bk^1_1}_f = \exp(-4.20\pm .0681) \f_\eff^{1.88 \pm .0229} \f^{.351 \pm .0300} \quad\text{for sphere+tube}.
\end{equation}
This is plotted in Figure \ref{fig:hybrid_fit}.  Some deviation persists at  low porosity, but it is less than an order of magnitude.  Both the sphere+tube data points and the tube data points collapse onto this curve.  \eqref{eq:hybrid_fit} is also consistent with \eqref{eq:tube_porosity}, the fit of porosity agaist permeability for the tube geometry.  Taking $\f_\eff =\f$ for the tubes, \eqref{eq:hybrid_fit} becomes
\begin{equation}
\label{eq:hybrid_fit_tubes}
\mean{\bk^1_1}_f = \exp(-4.20) \f^{2.23} 
\end{equation}
This is similar to the most general permeability relationships, formulated in \cite{scheidegger74pft, bear1988dfp}:
\begin{equation}
\textrm{permeability}  = \ell^2 f_1(\textrm{pore shape})f_2(\f)
\end{equation}
By including both $\f$ and $\f_\eff$ in \eqref{eq:hybrid_fit}, we capture some aspect of the pore shape.

\subsection{Bulk and Supplementary Viscosities Revisited}
\label{sec:visc_revisited}
In contrast to the permeability problem, the effective bulk viscosity and 
supplementary anisotropic viscosity are quite robust to the domain 
distortion. As before, we fit   $\mean{\zeta}_s$ to $(1-\f)^p/\f^q$, expecting 
$p$ and $q$ to be close to unity.  For the sphere+tube and the sphere 
geometry, the least squares fits are:
\begin{subequations}
\begin{align}
\label{eq:bulk_sphere_fit}
\mean{\zeta}_s &=\exp(0.301 \pm 0.0102 ) \f^{-1.00 \pm 0.00174}(1-\f)^{0.718 \pm 0.0337},\quad\text{for sphere geometry}\\
\label{eq:bulk_st_fit}
\mean{\zeta}_s &= \exp(0.124 \pm 0.0975) \f^{-0.985 \pm 0.0252}(1-\f)^{1.09 \pm 0.186},\quad\text{for sphere+tube geometry}
\end{align}
\end{subequations}
The data and these fits are plotted appear in Figure \ref{fig:bulkcomparison_plot}.  
The spherical geometry  appears to be an upper bound on the bulk viscosity for 
a given porosity.  We also remark that the prefactors vary by less than an order 
of magnitude amongst the different domains.  This is a strong endorsement of an 
effective bulk viscosity$\propto \f^{-1}$ not only for small porosity, but also for 
moderate porosities $\gtrsim 10\%$.

The supplementary anisotropic viscosity terms are also robust under this geometric 
perturbation.  For the problem $(l,m)=(1,1)$, the two new geometries fit:
\begin{subequations}
\begin{align}
\label{eq:e11_sphere}
-\mean{e_{1,1}(\bar{\chi}^{11})}_s &= \exp(-1.68 \pm .0588) \f^{.980 \pm .0101} (1-\f)^{3.56 \pm .195},\quad\text{for sphere geometry}\\
\label{eq:e11_st}
-\mean{e_{1,1}(\bar{\chi}^{11})}_s&= \exp(-1.94 \pm .139) \f^{.912 \pm .0359} (1-\f)^{1.25 \pm .265},\quad\text{for sphere+tube geometry}
\end{align}
\end{subequations}
The curves and the data are also plotted in Figure \ref{fig:e11comparison}.  For 
$\phi \lesssim 10\%$, the spread amongst the three geometries is less than an 
order of magnitude.

Similar results are found in the $(l,m)=(1,2)$ problem.  The two new additional 
domains are fit with:
\begin{subequations}
\begin{align}
\label{eq:e12_sphere}
-\mean{e_{12}(\bar{\chi}^{(12)})}_s &= \exp(-1.67 \pm .0222) \f^{1.02 \pm .000380}(1-\f)^{ 0.400 \pm .0737},\quad\text{for sphere geometry}\\
\label{eq:e12_st}
-\mean{e_{12}(\bar{\chi}^{(12)})}_s &= \exp(-1.32 \pm .00883)\f^{1.03\pm .00228}(1-\f)^{0.871 \pm .169},\quad\text{for sphere+tube geometry}
\end{align}
\end{subequations}
The sphere+tube data is bounded between the sphere data and the tube data; 
the spread is less than an order of magnitude.

\section{Discussion and Open Problems}
\label{sec:discuss}



We have successfully  parameterized the macroscopic parameters on
ensemble of domains for the simple pore geometries.  These can now be
consistently used with the macroscopic model given by equations
(\ref{eq:bi1} -- \ref{eq:bi3}).  We now review and discuss our
computations, both independently of and together with the macroscopic equations.

\subsection{Sensitivity to Geometry}
\label{sec:sensitivity}

As demonstrated by our computations in Sections \ref{sec:tubes} and \ref{sec:general_domains}, the effective parameters demonstrate a variety of sensitivities to the geometry.  Permeability, for our cell geometries, can be bounded by porosity, $k_\eff \lesssim \f^{n}$, where $n \sim 2$, as seen in Figure \ref{fig:perm-comparison}.  But in general, it cannot  be expressed as a function of a single parameter, such as porosity.  However, if the shapes are constrained by textural equilibration, as in \cite{cheadle1989pte}, there will be a single variable parameterization for each dihedral angle.  Solving the cell problems on these shapes, and assessing their sensitivity to the dihedral angle,  is an important open problem.

In contrast, the bulk viscosity appears to be rather insensitive to the geometry, scaling as $\zeta_\eff \propto \f^{-1}(1-\f)$.  We see this from the variation, or lack thereof, between the data and fits for the isolated spheres and the triply intersecting cylinders, in Figure \ref{fig:bulkcomparison_plot}.  Though these are entirely different geometric structures, there is less than an order of magnitude of variation in the computed $\zeta_\eff$.  These too merit computation on the texturally equilibrated shapes.  Randomly generated geometries may also be of interest.  

Both components of the supplementary anisotropic shear viscosity appear to be insensitive to geometry, with less than an order of magnitude amongst the three geometries.  However, because the problems are driven by surface stresses, it may be that a more anisotropic shape could alter these scalings.  Studying them on randomly generated shapes may provide insight on the role of grain scale anisotropy.

\subsection{Bulk Viscosity}

Perhaps our most significant result is the self-consistent bulk viscosity, arising from a \emph{purely mechanical model} of partially molten rock.   A spatially varying bulk viscosity is quite important.  Indeed, significant differences in dynamics were noted between the solutions of the \cite{mckenzie1984gac} model and the model in \cite{ricard2001b}.  The authors point to the use of a constant bulk viscosity in the McKenzie model as the source of the discrepancy.

Prior to \cite{ricard2001b}, \cite{schmeling2000pma} remarked that the $\f^{-1}$ dependence has an important impact on the compaction length.  For $\f=O(1\%)$, the bulk viscosity is two orders of magnitude greater than the shear viscosity.  While many studies took $\zeta_s$ and $\mu_s$ to be the same order, this higher bulk viscosity leads to a compaction length an order of magnitude greater.  \cite{ricard2007pmc} made a similar observation on the impact of variable bulk viscosity on the compaction length.

Schmeling also commented that this variable bulk viscosity could
induce melt focusing towards the axis in his plume simulations.  This
is an additional nonlinearity that may be important to geophysical
problems.  Many studies relying on the McKenzie model employed a
constant bulk viscosity, including \cite{richter1984dmm,
  spiegelman1987sdm, Spiegelman1993a, Spiegelman1993b,
  spiegelman1993pme, aharonov1997tdf, spiegelman1996gcm,
  kelemen1997rmm, spiegelman2001cac, katz2004rma, spiegelman07aia}.
\cite{spiegelman2003ecv,spiegelman2003lam} used a bulk viscosity, with
$\zeta_s \propto \f^{-m}$ with $m>n$, $n$ the exponent in the
permeability relationship $\kappa \propto \f^n$ to prevent the system
from compacting to zero between the reactive channels.  It would be
interesting to revisit these problems with a $\f^{-1}$ bulk viscosity.


\subsection{Compaction Length}
\label{sec:compaction}

We now combine (\ref{eq:bi1} -- \ref{eq:bi3}), our leading order equations derived by multiple scale expansions, with our computed constitutive relations.  For $\f \ll1$ our numerical estimates, \eqref{eq:hybrid_fit}, (\ref{eq:bulk_sphere_fit}--\ref{eq:bulk_st_fit}), (\ref{eq:e11_tube}--\ref{eq:e11_st}), \eqref{eq:e11_sphere}, and (\ref{eq:e12_tube}--\ref{eq:e12_sphere}) are approximately:
\begin{align*}
\zeta_\eff &\approx \mu_s\zeta_0 \f^{-1}(1-\f)\\
\eta_\eff & \approx  \mu_s\eta_0 \f (1-\f) \\
k_\eff & \approx k_0 \ell^2\f_\eff^{1.9}\f^{.35}
\end{align*}
$k_0$ is a $O(10^{-3}-10^{-2})$ constant, $\zeta_0$ is an $O(1)$ constant, and $\eta_0$ is a $O(10^{-1})$ constant fourth order tensor.  Under these assumptions, the equations for the Biphasic-I model, (\ref{eq:bi1} -- \ref{eq:bi3}), simplify:
\begin{gather}
\begin{split}
0&=\overline{\rho}\bg-\nabla P +\nabla\bracket{\mu_s\zeta_0 \f^{-1} \nabla \cdot \bV^s}\\
&\quad+ \nabla \cdot\bracket{2(1-\f){\mu_s }e(\bV^s)   -\frac{2}{3}(1-\f)\mu_s \nabla \cdot \bV^s I}\\
&\quad + \nabla \cdot \bracket{2\mu_s \f (1-\f) \eta_0^{lm} e_{x,lm}(\bV^s)}
\end{split}\\
\phi(\bV^f - \bV^s) = - \frac{k_0\ell^2 \f_\eff^{1.9} \f^{.35}}{\mu_f}\paren{\nabla P- \bg^f}\\
\nabla \cdot\bracket{\f{\bV^f} + (1-\f)\bV^s}=0
\end{gather}

If we were to use these equations and numerically derived constitutive relations to solve a boundary value problem, the compaction length would again appear as an important length scale,
\[
\begin{split}
\delta_\textrm{comp.} &= \sqrt{\frac{\bracket{\zeta_\eff + \frac{4}{3}\mu_s(1-\f) + 2 \abs{\eta_\eff}} k_\eff }{\mu_f}}\\
&\approx \ell\sqrt{ \frac{\mu_s (1-\f)(\zeta_0 \f^{-1}+\frac{4}{3} + 2 \abs{\eta_0}\f) k_0 \f_\eff^{1.9}\f^{.35}}{\mu_f}}
\end{split}
\]
If $\f \ll1$, then  $\zeta_0\f^{-1} \gg 4/3 + 2 \abs{\eta_0}\f$ and $1-\f\approx 1$,
\begin{equation}
\label{eq:compaction_length_est}
\delta_\textrm{comp.} \approx \ell\sqrt{\zeta_0 k_0\mu_s/\mu_f}  \f^{-.325} \f_\eff^{.95}
\end{equation}
Since $\f_\eff \leq \f$, we have an upper bound on the compaction length,
\begin{equation}
\label{eq:compaction_length_bound}
\boxed{\delta_\textrm{comp.} \lesssim \ell\sqrt{\zeta_0 k_0\mu_s/\mu_f} \f^{.6}}
\end{equation}
Hence,
\[
\lim_{\f\to 0} \delta_\textrm{comp.} = 0
\]
We believe \eqref{eq:compaction_length_bound}, which constrains the compaction length by the porosity raised to a small positive power, is relatively insensitive to the geometric configuration.  This follows from our alleged robustness of our effective bulk viscosity, $\zeta_\eff \propto \f^{-1}$, and the broad agreement in the porosity--permeability relationship, $\kappa \propto \f^n$ with $n\geq 2$.

A compaction length that vanishes with porosity has interesting consequences.  For example, this compaction length scaling does not rule out the possibility that a partially molten rock could expel all fluid by mechanical means.  It also does not permit the infiltration of fluid into a dry region.  Understanding how any of these systems of equations transition between a partially molten region and a dry region is an outstanding question.  We also note that though our scaling relationship does not forbid compaction, it does not imply it either.  There may be a dynamic response that prevents the matrix from mechanically compacting to zero.  Such effects were mathematically proven to exist in a one-dimensional simplification of the model, without melting or freezing, in \cite{Simpson07,simpson08as,simpson08hpde}.

If, instead,  we had concluded $\delta_\textrm{comp.} \propto \f^{q}$, with $q<0$, then the compaction length would become unbounded as the melt vanished.  Hence the region of deformation in the matrix needed to segregate additional fluid would also become infinite, precluding further segregation solely by mechanical processes.

Because we have parameterized the permeability with $\f_\eff$, we can explicitly see the response of the  compaction length as the melt network becomes disconnected.  $\f_\eff$ measures the volume fraction where melt flows.  As the channels close up and the melt becomes trapped and $\f_\eff \to 0$.  Taking this limit in \eqref{eq:compaction_length_est},
\[
\lim_{\f_\eff \to 0} \delta_\textrm{comp.} = 0
\]
The compaction length can vanish, \emph{even if the melt fraction remains bounded away from zero}.  A similar conclusion could be drawn for the compaction length of McKenzie,
\begin{equation}
\delta_{\textrm{M84}} = \sqrt{\frac{\kappa(1-\f)(\zeta_s + \frac{4}{3} \mu_s)}{\mu_f}}
\end{equation}
Letting $\kappa = \kappa_0\f^n$, if we interpret the loss of connectivity as $\kappa_0 \to 0$, $\delta_{\textrm{M84}}$ vanishes with nonzero porosity.  

\subsection{Other Physics}
\label{sec:other_phsysics}

There are several results that were not realized by our model, and
these merit discussion.  We did not recover the $\zeta_s \propto
\log(\f^{-1})$ result of \cite{arzt1983pah}.  This arises in the limit
of sufficiently low porosity that the primary transport mechanism is
by grain boundary diffusion.  Since we did not include any surface
physics in our fine scale model, we should not have expected to
upscale their effects.  If good grain scale descriptions of these
processes could be formulated, it might be possible to coarsen them
via homogenization, perhaps  recovering this macroscopic
relation. However, as  $\log(\f^{-1})$ becomes unbounded more slowly
than $\f^{-1}$, this result does not change the basic argument about
the compaction length vanishing with zero porosity.

Another relationship not captured by either our work is the 
experimental fit for matrix shear viscosity in the presence of melt 
from \cite{hirth1995ecd1, hirth1995ecd2, kelemen1997rmm,kohlstedt2000rpm, kohlstedt2007prm},
\[
\mu_{s+f} \propto \exp \paren{-\f/\f_\ast}
\]
Our model possesses a porosity weakening mechanism; all of the 
viscosity terms are $\propto 1-\f$.  However, the anisotropic part, 
$\eta_\eff$, is not sign definite and is small compared to the isotropic 
component.  Furthermore, there does not appear to be an exponential 
relation.  \cite{hirth1995ecd1} hypothesized that the presence of melt 
enhances grain boundary diffusion, providing a fast path for 
deformation through the melt.  As with the $\log(\f^{-1})$ bulk viscosity, 
this is a surface physics phenomenon not captured by our Stokes models.

\subsection{Open Problems}

As discussed, two important open problems are the computation of 
the cell problems on more realistic and general cell domains and the 
inclusion of surface physics into the model.  The former problem is 
rather straightforward, requiring good computational tools for 
generating the domains and solving the Stokes equations on them.  
The latter problem is more challenging, requiring fine scale equations 
for these processes.  

The models could also be augmented by giving the matrix a nonlinear
rheology, leading to nonlinear cell problems.  Though these could be
solved and studied numerically, this is much more difficult as the
different forcing components no longer decouple.  Rather than being
able to split the matrix cell problems into a bulk viscosity problem,
and two surface stress problems, they would have to be done
simultaneously.   However, the derived parameterizations for the
effective viscosities would be important to magma migration.  A
nonlinear matrix rheology is expected at large strain rates and  was 
needed to computationally model physical experiments for shear 
bands in \cite{katz2006dma}.

Another important physical rheology is viscoelasticity.
\cite{connolly1998compaction, vasilyev1998modeling} extended the
earlier, purely viscous, models to include elastic effects.  This is
an important regime since it permits both short and long time scales,
as is found at the asthenosphere-lithosphere boundary.

%


%
%
%
%
%
%

%
%
%
%
\appendix

\section{Spherical Model}
\label{sec:bulk_sphere}


Here we develop a toy model for equations (\ref{eq:dilation1} -- \ref{eq:dilation3}),
\begin{align*}
\nabla_{y}\cdot\paren{-\zeta I + 2 e_{y}(\bar{\xi})}&=0\quad\text{in $Y_s$}\\
\nabla_y \cdot \bar{\xi}&= 1\quad \text{in $Y_s$}\\
\paren{-\zeta I + 2 e_{y}(\bar{\xi})}\cdot \bn& =0\quad \text{on $\gamma$}
\end{align*}
whose solution yields the effective bulk viscosity,
\[
{\zeta_\eff =\mu_s \mean{\zeta}_s-\frac{2}{3}\mu_s (1-\f)}
\]
Consider a fluid domain, $Y_f$, occupying a small isolated sphere at the center of the unit cube; $Y_s$ is the complementary region.  Smoothing out the exterior boundary of $Y_s$ deforms it into a sphere.  We solve the dilation stress problem on this domain.  To avoid confusion, let
\begin{align}
{Y_s}^\textrm{sphere} &= \set{\by\in\R^3 \mid a\leq\abs{\by}\leq 1 }\\
{Y_s}^\textrm{cube}&= \set{\by\in\bracket{-\frac{1}{2},\frac{1}{2}}^3\mid \abs{\by}\geq a}
\end{align}
Our toy problem is posed on ${Y_s}^\textrm{sphere}$.


Although periodicity is no longer a meaningful boundary condition, it can be shown that the normal velocity on the periodic part of $\partial Y_s$ vanishes.  We set the normal velocity to zero on the exterior boundary of ${Y_s}^\textrm{sphere}$.  On the interior shell, the stress free condition remains.  The equations are:
\begin{subequations}
\begin{align}
\label{eq:bulk_cell1}
\nabla \cdot \paren{- \zeta I + 2 e(\bar{\xi})} &=0\quad\text{in ${Y_s}^\textrm{sphere}$}\\
\label{eq:bulk_cell2}
\nabla \cdot \bar{\xi} &=1\quad\text{in ${Y_s}^\textrm{sphere}$}\\
\label{eq:bulk_cell3}
\paren{- \zeta I + 2 e(\bar{\xi})} \cdot \bn&=0\quad\text{at $r=a$}\\
\label{eq:bulk_cell4}
\bar{\xi}\cdot \bn &= 0\quad\text{at $r=1$}
\end{align}
\end{subequations}
$a<1$ is the radius of the interior sphere.  Decomposing the velocity into incompressible, $\bv^\inc$, and compressible, ${\nabla \Pi}$, components, the compressible part solves:
\begin{equation}
\nabla^2 \Pi = 1
\end{equation}
Let the boundary conditions on the potential be:
\begin{equation}
\Pi |_{r=a}  =0, \quad \nabla \Pi |_{r=1} =0
\end{equation}
Since the problem is spherically symmetric, its solution is
\begin{equation}
\Pi = \frac{1}{6}(r^2-a^2) + \frac{1}{3}(r^{-1}-a^{-1})
\end{equation}

The incompressible velocity must be divergence free.  Again, by spherical symmetry,
\[
\nabla \cdot \bv^\inc = \frac{1}{r^2}\partial_r \paren{r^2 v_r^\inc}= 0 \Rightarrow v_r^\inc = C/r^2
\]
To satisfy the boundary condition at $r=1$, $\bv^\inc \cdot \bn = v_r^\inc =0$.  Therefore, $C=0$ and $\bv^\inc = 0$.  The pressure then solves $\partial_r \zeta =0$, so it is constant,
\begin{equation*}
\zeta(r)= \zeta(a)\quad\text{for $r \in (a,1)$}
\end{equation*}

Applying boundary condition \eqref{eq:bulk_cell3},
\begin{equation*}
\zeta = 2 e_{rr}(\bar{\xi})|_{r=a}=  2 \partial_r^2 \Pi|_{r=a}= \frac{1}{3} \paren{2 + \frac{4}{a^3}}
\end{equation*}

In this geometry, $\f = a^3$, and the cell averaged pressure is
\begin{equation}
\label{eq:meanzeta_sphere}
\mean{\zeta}_s = \frac{4}{ 3 \phi}\paren{ 1 + \frac{\f}{2}}\paren{1-\f}
\end{equation}
Therefore,
\begin{equation}
\label{eq:bulk_eff_sphere}
\zeta_\eff= \mu_s\mean{\zeta}_s - \frac{2}{3}\mu_s(1-\f) = \frac{4 \mu_s}{3 \f}\paren{1 -\f}
\end{equation}
\eqref{eq:bulk_eff_sphere} uses definition \eqref{eq:bulk_eff}, but the $\zeta$'s in 
each solve different problems.  \eqref{eq:bulk_eff_sphere} is specific to the geometry 
of a sphere with cavity, while \eqref{eq:bulk_eff} is for a generic geometry occupying 
some fraction of the unit cube.

\begin{figure*}[t]
\noindent\includegraphics{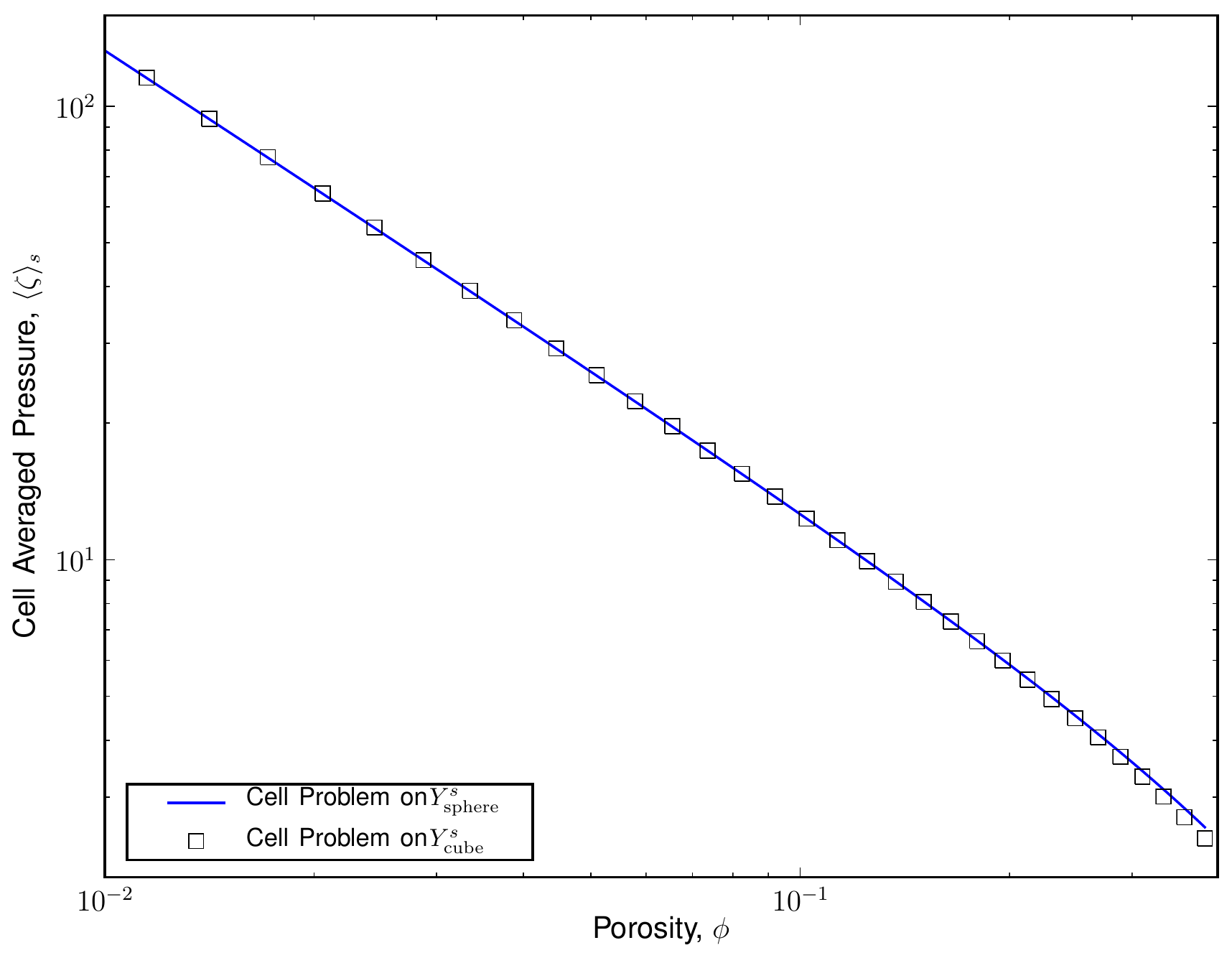}
\caption{Cell averaged pressure, \eqref{eq:meanzeta_sphere}, plotted as a function of porosity.   The data from the numerical solution of the cell problem given by equations (\ref{eq:dilation1} -- \ref{eq:dilation3}) and posed on ${Y_s}^\textrm{cube}$ also appears.  It is in good agreement with the analytic solution on the spherical domain, ${Y_s}^\textrm{sphere}$.}
\label{fig:bulk_spheres}
\end{figure*}

\eqref{eq:meanzeta_sphere} is plotted along with data for the numerical solutions
 of the cell problem posed on ${Y_s}^\textrm{cube}$ in Figure \ref{fig:bulk_spheres}. 
 There is good agreement between \eqref{eq:meanzeta_sphere} and these computations 
 for porosity $\gtrsim 10\%$ suggesting our deformation 
 ${Y_s}^\textrm{cube} \Rightarrow {Y_s}^\textrm{sphere}$ was reasonable.  We reiterate 
 that though our result is quite similar to that of \eqref{eq:bulkviscosity_taylor} and 
 \eqref{eq:bulkviscosity_prudhomme}, the origin of the underlying Stokes problem is quite different.

\section{Computational Methods and Results}
\label{sec:computation_notes}

\subsection{Cell Problem Boundary Conditions}
\begin{figure*}[p]
\centering
\includegraphics[width=6in]{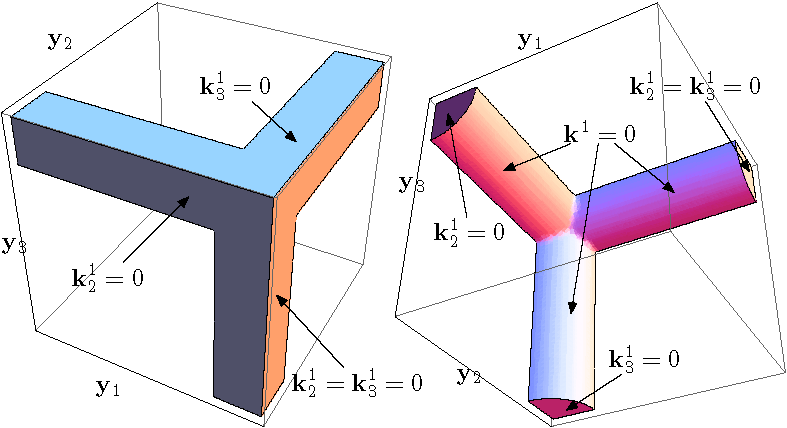}
\caption{Specification of Dirichlet boundary conditions on $\mathbf{k}^1$ when solving on a symmetry reduced domain.}
\label{fig:perm_bcs}
\end{figure*}

Though we could solve the cell problems as stated, with the specified boundary conditions on interface $\gamma$ and periodic on the rest of the domain, we use the symmetry of our model problems to reduce the computational cost by a factor of eight.  The symmetry properties of the solutions allow us to formulate the appropriate boundary conditions on the portion of the boundary that is not $\gamma$.    The symmetry properties of the cell problems are summarized in Table \ref{table:symmetry}.  We specify these Dirichlet boundary conditions on the velocity together with the original boundary condition on $\gamma$.  When forming the weak form of the problem, we use neutral boundary conditions, $\sigma \cdot \bn = 0$ on the part of the boundary that is not $\gamma$.  For example, in solving the permeability cell problem of Section \ref{sec:perm} for $\mathbf{k}^1$ and $q_1$, the weak form is
\begin{equation}
\int_{Y_f} \nabla \mathbf{k}^1 : \nabla \bar{\phi} - q_1 \nabla \cdot \bar{\phi} - \psi \nabla \cdot \mathbf{k}^1 = \int_{Y_f} \bar{\phi} \cdot \mathbf{e}_1
\end{equation}
where $\bar{\phi}$ and $\psi$ are test functions.  The Dirichlet boundary conditions are indicated in Figure \ref{fig:perm_bcs}.  On the part of the boundary that is the interface, $\gamma$, we have applied the no - slip condition.  

\begin{table}[h]
\centering
\caption{Summary of symmetry conditions used as Dirichlet boundary conditions in the cell problem computations.}
\label{table:symmetry}
\begin{tabular}{r|ccc}
Cell Problem Velocity & $y_1=0, -.5$ & $y_2=0,-.5$ & $y_3=0,-.5$\\
\hline
$\mathbf{k}^1$& $\mathbf{k}^1_2=\mathbf{k}^1_3=0$& $\mathbf{k}^1_2=0$&$\mathbf{k}^1_3=0$\\
$\bar{\chi}^{11}$&$\bar{\chi}_1^{11}=0$ &$\bar{\chi}_2^{11}=0$ &$\bar{\chi}_3^{11}=0$\\
$\bar{\chi}^{12}$&$\bar{\chi}_2^{12},\bar{\chi}_3^{12}=0$ &$\bar{\chi}_1^{12}=\bar{\chi}_3^{12}=0$ &$\bar{\chi}_3^{12}=0$\\
$\bar{\xi}$ & $\bar{\xi}_1=0$& $\bar{\xi}_2=0$&$\bar{\xi}_3=0$
\end{tabular}
\end{table}

\subsection{Solver Algorithms}
\label{sec:numerical_methods}
We discretize the Stokes equations for the cell problems using the
P2-P1 formulation described in \cite{elman2005fea}.  The FEniCS
libraries are used to generate code for the weak forms of the
equations and assemble the associated matrices and vectors,
\cite[]{DupHof2003,KirLog2006,KirLog2007,Log2007}.  These vectors and
matrices are passed to PETSc and solved using   algebraic Multigrid preconditioned GMRES, \cite[]{petsc-efficient,petsc-user-ref,petsc-web-page}.  Domains and meshes were created with CUBIT \cite[]{cubit}. The versions of the software we used are summarized in Table \ref{table:software}.

\begin{table}
\centering
\caption{Software versions}
\label{table:software}
\begin{tabular}{r|r}
Package & Version\\
\hline
CUBIT&11.0\\
DOLFIN(FEniCS)& 0.7.2\\
FFC(FEniCS)&0.4.4\\
FIAT(FEniCS)&0.3.4\\
HYPRE&2.0.0\\
PETSc& 2.3.3\\
UFC(FEniCS)&1.1\\
UMFPACK&4.3
\end{tabular}
\end{table}

To study problems with $O(10,000-100,000)$ elements and $O(100,000-1,000,000)$ unknowns, we rely on a Stokes preconditioner employing the {pressure mass matrix} of \cite{elman2005fea}.  The Stokes system is
\[
\begin{pmatrix}\bold{A}& \bold{B}^T\\ \bold{B}& \bold{0} \end{pmatrix}\begin{bmatrix}\bold{u}\\ \bold{p} \end{bmatrix}= \bold{K} \begin{bmatrix}\bold{u}\\ \bold{p} \end{bmatrix}= \begin{bmatrix}\bold{f} \\ \bold{g}\end{bmatrix}
\]
where $\mathbf{A}$ is matrix corresponding to the weak form of the vector Laplacian, $\mathbf{B}^T$ is the matrix corresponding to the weak form of the gradient, and $\mathbf{B}$ is the matrix corresponding to the weak form of the divergence.  This is preconditioned with an approximate inverse of 
\[
\bold{P}=\begin{pmatrix}\bold{A}& \bold{0}\\ \bold{0}& \bold{Q} \end{pmatrix}
\]
$\bold{Q}$ is the {pressure-mass matrix}.  

As our meshes are unstructured, the HYPRE library is used for
algebraic multigrid preconditioning.  In particular, we use BoomerAMG.
We apply this on all of $\bold{P}$, although we could have only used
this on the $\bold{A}$ block, and relied on Jacobi or another light weight
pre-conditioner for the $\bold{Q}$ block.  

\subsection{Examples and Benchmarks}
As a test, we solve the permeability cell problem of Section \ref{sec:perm}, with the symmetry reductions, for flow past a sphere of radius $0.3$.  It is meshed with a characteristic size of $.03125$, consisting of 119317 tetrahedrons.  The results are summarized in Table \ref{table:convergencecomparison}

\begin{table}
\centering
\caption{Convergence comparison between solvers}
\label{table:convergencecomparison}
\begin{tabular}{r|l}
Method & $\mean{\mathbf{k}_1^1}$\\
\hline
BoomerAMG on $\bold{P}$ + GMRES & 0.0447051 \\
JT05  & 0.045803\\
COMSOL & 0.044497
\end{tabular}
\end{table}

For comparison, \cite{jung2005fpt} ran a time dependent problem to steady state and used an immersed boundary method with finite volumes. In our COMSOL computation, we used ``fine" meshing, with 29649 elements, 134260 degrees of freedom, and a relative tolerance of 1e-10 in the solver.


The objective function, $\mean{\mathbf{k}_1^1}$, converges as we refine our mesh; see Table \ref{table:convergence1} for a comparison of different meshes for this problem.

\begin{table}
\centering
\caption{Convergence data for permeability cell problem I}
\label{table:convergence1}
\begin{tabular}{r|rrll}
Mesh Size & No. Cells & No. d.o.f. & $\mean{\mathbf{k}_1^1}_f$& $\abs{\% \Delta}$\\
\hline
0.25 & 56 & 430 & 0.0492875& --\\
0.125 & 534 & 2950 & 0.0472207&0.0419336\\
0.0625 & 3673&17988&0.0452142&0.042492 \\
0.03125 & 26147&119317&0.0447051& 0.0112597\\
auto &14803&70525&0.0445419 &--
\end{tabular}
\end{table}


Table \ref{table:convergence2} summarizes the convergence results for flow around a sphere of radius $0.45$.  Again, our method appears to be quite effective.

\begin{table}[h]
\centering
\caption{Convergence data for permeability cell problem II}
\label{table:convergence2}
\begin{tabular}{r|rrll}
Mesh Size & No. Cells & No. d.o.f. & $\mean{\mathbf{k}_1^1}_f$& $\abs{\% \Delta}$\\
\hline
0.25 & 61 & 475 & 0.00763404& -- \\
0.125 & 384 & 2302 & 0.00651896&0.146067 \\
0.0625 & 2620&13370&0.00626809&0.0384831 \\
0.03125 & 19916&93011&0.00617889&0.0142308 \\
auto &23776&112620&0.00616139&-- \\
JT05& -- & --&0.0064803&--\\
COMSOL&--&--&0.006153&--\\
\end{tabular}
\end{table}

As another example, we solve the dilation stress cell problem from Section \ref{sec:bulk} .  Solved on the domain complementary to a spherical inclusion of radius $0.2$, the convergence results are summarized in Table \ref{table:convergence3}.

\begin{table}
\centering
\caption{Convergence data for dilation stress cell problem}
\label{table:convergence3}
\begin{tabular}{r|rrll}
Mesh Size & No. Cells & No. d.o.f. & $\mean{\zeta}_s$& $\abs{\% \Delta}$\\
\hline
0.25 & 73 & 541 & 46.6432&-- \\
0.125 & 514 & 2862 & 44.1747&0.052923\\
0.0625 & 3813&18575&40.7177&0.0782575 \\
0.03125 & 29063&132115&39.4805&0.0303848 \\
auto &13725&64205&39.1558&-- \\
COMSOL&--&--&39.117074&--
\end{tabular}
\end{table}

\begin{figure*}[p]
\centering
$\begin{array}{cc}
\includegraphics[width=2.5in]{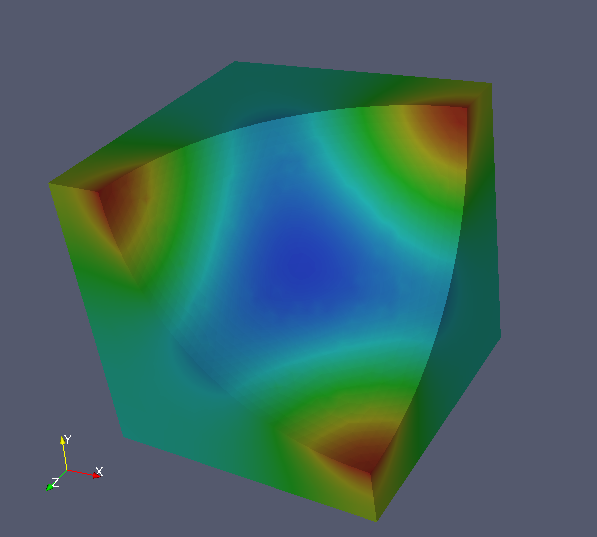}&\includegraphics[width= 2.5in]{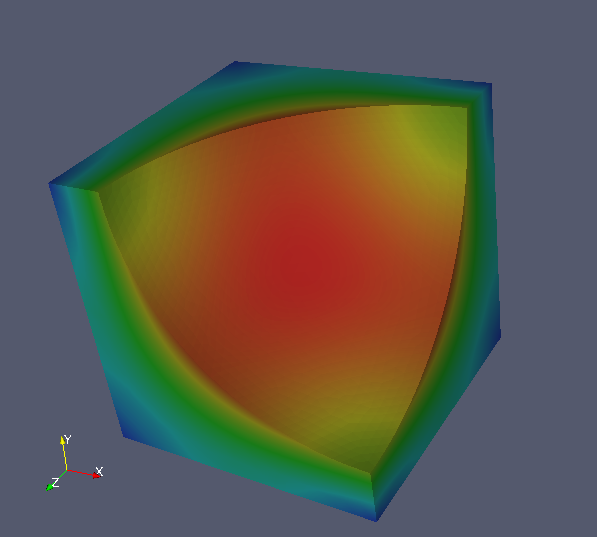}\\
\text{\textbf{(a)}}& \text{\textbf{(b)}}\\
\includegraphics[width= 2.5in]{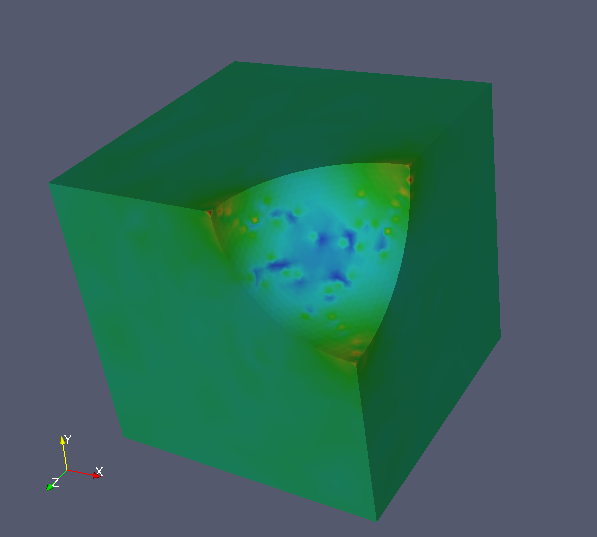}&\includegraphics[width= 2.5in]{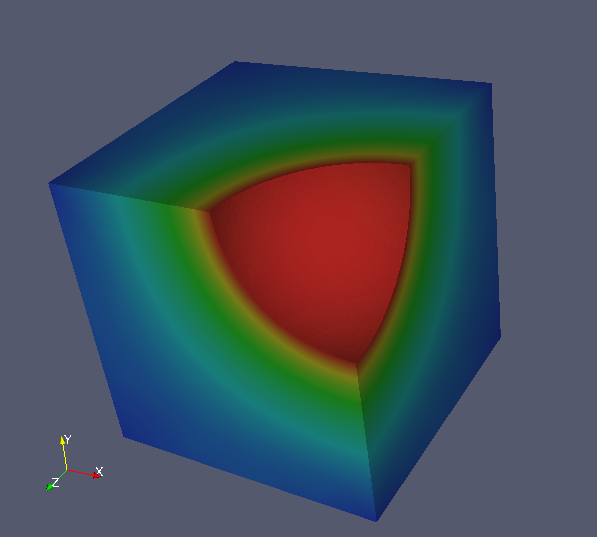}\\
\text{\textbf{(c)}}& \text{\textbf{(d)}}\\
\includegraphics[width= 2.5in]{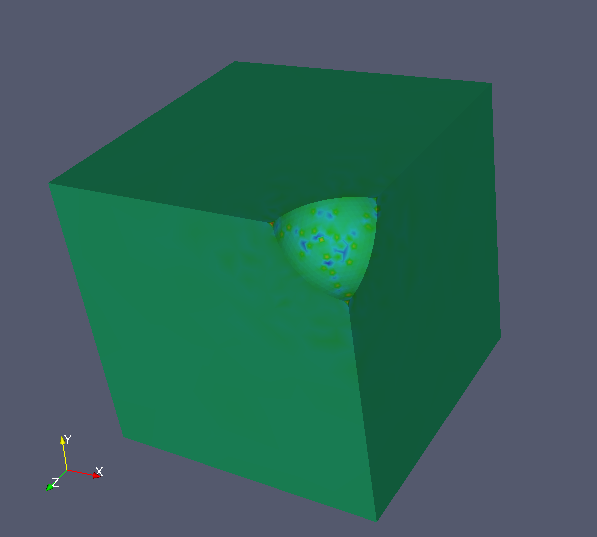}&\includegraphics[width= 2.5in]{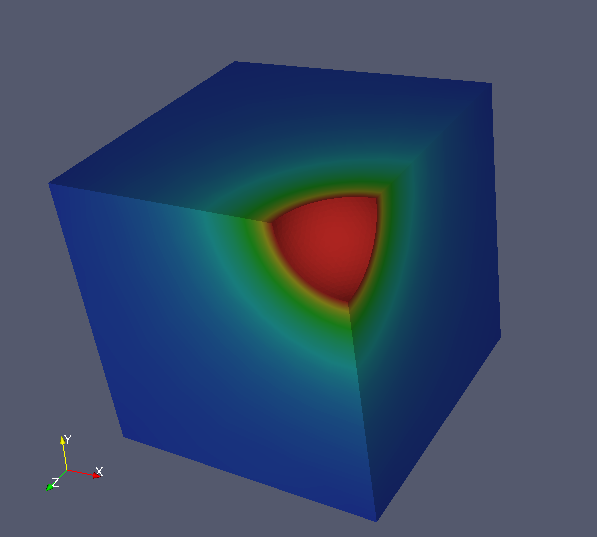}\\
\text{\textbf{(e)}}& \text{\textbf{(f)}}
\end{array}$
\caption{Figures on the left are the pressure fields for domains complementing spheres of radii $a = .40$, $a=.20$, and $a=.10$.  Figures on the right are the corresponding velocity magnitude fields.}
\label{fig:pressure_convergence}
\end{figure*}

The data in Tables \ref{table:convergence1}-- \ref{table:convergence3} were computed with default PETSc KSP tolerances.  The automatically generated mesh was constructed with  the CUBIT command:
\begin{verbatim}
volume 6 sizing function type skeleton scale 3 time_accuracy_level 2
min_size auto max_size 0.2 max_gradient 1.3
\end{verbatim}

While these convergence results are encouraging, our data is imperfect.  Continuing with the dilation stress example, consider the data in Figure \ref{fig:pressure_convergence}.  Comparing Figures \textbf{(a)}, \textbf{(c)}, and \textbf{(e)}, it would appear that the domains with smaller fluid inclusions have less well resolved pressure fields.  They could likely be resolved with additional resolution.  However, we use this data and believe it to be valid for several reasons:
\begin{enumerate}
\item It is preferable to have all domains meshed with the same algorithm.
\item While the pressure fields may not be resolved, the error appears at the interface, and we are interested in the cell average.  Moreover, the relative variations about the cell average are small.
\item The corresponding velocity fields, with magnitudes pictured in Figures \textbf{(b)}, \textbf{(d)}, and \textbf{(f)}, appear to be smooth, suggesting we are converging towards the analytical solution.
\item The cell averages are consistent with the trends from the better resolved cases.
\end{enumerate}

\subsection{Cell Problem Data}
\label{sec:celldata}
All meshes were generated using CUBIT with the command:
\begin{verbatim}
volume 6 sizing function type skeleton scale 3 time_accuracy_level 2
min_size auto max_size 0.2 max_gradient 1.3
\end{verbatim}
Problems were solved in PETSc with a relative tolerance of $10^{-8}$ and and absolute tolerance of $10^{-50}$.






\begin{acknowledgments}
Both this paper and \cite{simpson08a} are based on the thesis of G.~Simpson, \cite{simpson08t}, completed in partial fulfillment of the requirements for the degree of doctor of philosophy at Columbia University.

The authors wish to thank  D.~Bercovici and R.~Kohn for their helpful comments.

This work was funded in part by the US National Science Foundation (NSF) Collaboration in Mathematical Geosciences (CMG), Division of Mathematical Sciences (DMS), Grant DMS--05--30853, the NSF Integrative Graduate Education and Research Traineeship (IGERT) Grant DGE--02--21041, NSF Grants DMS--04--12305 and DMS--07--07850.
\end{acknowledgments}

%
%

\bibliography{homogenization_agu_pt2_alt}

%
%
%
%
%
%
%
%


%
%

\end{article}





%
%

\end{document}